\documentclass[aps,prb,twocolumn,superscriptaddress]{revtex4-1}
\usepackage{amsmath,amsthm,amssymb}
\usepackage[dvipdfmx]{graphicx}
\usepackage{amsmath,bm}
\usepackage{color,ulem}
\usepackage{ifthen}

\newcount \commentout
\newcommand{\1}{\mbox{1}\hspace{-0.25em}\mbox{l}}

\newlength{\figwidth}
\newlength{\figlarge}
\begin{document}
%%%%%%%%%%%%%%%%%%%%%%%%%%%%%%%%%%%%%%%%%%%%%%%%%%%%%%%%%%%%%%%%%%%%%%%
\title{
Efficient method to compute $\mathbb{Z}_4$-indices with glide symmetry and applications to M\"obius materials, $\mathrm{CeNiSn}$ and $\mathrm{UCoGe}$
}
%%%%%%%%%%%%%%%%%%%%%%%%%%%%%%%%%%%%%%%%%%%%%%%%%%%%%%%%%%%%%%%%%%%%%%%
\author{Tsuneya Yoshida}
%\affiliation{Department of Physics, Kyoto University, Kyoto 606-8502, Japan}
\affiliation{Department of Physics, University of Tsukuba, Ibaraki 305-8571, Japan}
\author{Akito Daido}
\affiliation{Department of Physics, Kyoto University, Kyoto 606-8502, Japan}
\author{Norio Kawakami}
\affiliation{Department of Physics, Kyoto University, Kyoto 606-8502, Japan}
\author{Youichi Yanase}
\affiliation{Department of Physics, Kyoto University, Kyoto 606-8502, Japan}
%%%%%%%%%%%%%%%%%%%%
%%%%%%%%%%%%%%%%%%%%%%%%%%%%%%%%%%%%%%%%%%%%%%%%%%%%%%%%%%%%%%%%%%%%%%%
\date{\today}
%%%%%%%%%%%%%%%%%%%%%%%%%%%%%%%%%%%%%%%%%%%%%%%%%%%%%%%%%%%%%%%%%%%%%%%
\begin{abstract}
We propose an efficient method to numerically evaluate $\mathbb{Z}_4$-indices of M\"obius/hourglass topological phases with glide symmetry. 
Our method directly provides $\mathbb{Z}_4$-indices in the lattice Brillouin zone while the existing method requires 
careful observations of momentum dependent Wannier charge centers.
As applications, we perform systematic computation of $\mathbb{Z}_4$-indices for M\"obius materials, $\mathrm{CeNiSn}$ and $\mathrm{UCoGe}$. 
In particular, our analysis elucidates that $\mathrm{UCoGe}$ shows strong M\"obius superconductivity for the $A_u$- or $B_{3u}$-representation whose topology has not been fully characterized.
Furthermore, obtained phase diagrams reveal novel topological gapless excitations in the bulk which are protected by nonsymmorphic glide symmetry.
We observe these gapless excitations with glide symmetry by doping holes into the superconducting phase of the $B_{1u}$- or $B_{3u}$-representation.
\end{abstract}
%%%%%%%%%%%%%%%%%%%%%%%%%%%%%%%%%%%%%%%%%%%%%%%%%%%%%%%%%%%%%%%%%%%%%%%
\pacs{
***
} 
%%%%%%%%%%%%%%%%%%%%%%%%%%%%%%%%%%%%%%%%%%%%%%%%%%%%%%%%%%%%%%%%%%%%%%%
%%%%%%%%%%%%%%%%%%%%%%%%%%%%%%%%%%%%%%%%%%%%%%%%%%%%%%%%%%%%%%%%%%%%%%%
\maketitle
%%%%%%%%%%%%%%%%%%%%%%%%%%%%%%%%%%%%%%%%%%%%%%%%%%%%%%%%%%%%%%%%%%%%%%%

%%%%%%%%%%%%%%%%%%%%%%%%
%\input{intro.tex}
%%%%%%%%%%%%%%%%%%%%%%%%
%%%%%%%%%%%%%%%%%%%%%%%%
\section{Introduction}
%%%%%%%%%%%%%%%%%%%%%%%%

In the last decade, topological perspective on condensed matter systems has been rapidly developed along with the material realization of topological insulators and superconductors~\cite{TI_review_Hasan10,TI_review_Qi10}.
The realization is important not only to deepen our insights but also to develop technology. 
For instance, dissipationless spin current is expected to be applied to spintronic devices. 
Furthermore, Majorana fermions of topological superconductors attract much interest in terms of quantum computation~\cite{Kitaev_chain_01,Alicia_IOP12,Sato_JPSJ17}. 
So far, many efforts have been made to realize various topological insulators/superconductors~\cite{exp_2D-QW_MKonig_2007,exp_3D-bismuth_YXia_2008,exp_3D-bismuth_YLChen,exp_3D_bismuth_Se_Zhang,Mourik_Majorana2012,Majorana_Rokhinson2012,Majorana_Das2012,Nakosai_PRL12,exp-3D_TSC_Sasaki11} listed in classification tables~\cite{Schnyder_classification_free_2008,Kitaev_classification_free_2009,Ryu_classification_free_2010} of a so-called ten-fold way which is obtained by focusing on time-reveral, particle-hole, and chiral symmetry.

Remarkably, the notion of topological phases has been extended to topological crystalline insulators/superconductors~\cite{Fu_TCI_2011,Hsieh_TCH_SnTe_2012,Tanaka_TCI_SnTe2012,superlattice_Yanase_12,Weng_Dai_TCIinYbB12_2014,Yoshida_ZtoZ8_PRL17,TCI_Kruthoff_PRX17}. 
It has turned out that large variety of topological phases exist in materials thanks to the complexity and diversity of their crystal structure. Among them, topological crystalline phases with nonsymmorphic symmetry show intriguing behaviors~\cite{Shiozaki_Moebius_class_PRB16,Wang_hourglass_Nat2016}. 
Because of the symmetry transformation accompanied by the half-translation, these phases host M\"obius/hourglass surface states. Furthermore, this peculiar surface state results in $\mathbb{Z}_4$-classification in the presence of the time-reversal symmetry while topological phases with the local symmetry follow $\mathbb{Z}$- or $\mathbb{Z}_2$-classification. This remarkable discovery urges us to search a new series of topological insulators/superconductors.
After the proposal of M\"obius topological insulators/superconductors, realization of a $\mathbb{Z}_4$-M\"obius insulator with glide symmetry is reported for $\mathrm{KHgX}$ ($\mathrm{X}=$ $\mathrm{As}$, $\mathrm{Sb}$, $\mathrm{Bi}$)~\cite{Wang_hourglass_Nat2016,Ma_hourglass_exp_17}. 
In addition, the possibility of $\mathbb{Z}_4$-topological insulators/superconductors with glide symmetry is discussed for heavy-fermion materials; possibility of a M\"obius Kondo insulator is discussed for $\mathrm{CeNiSn}$~\cite{Chang_Moebius_CeNiSn_NatPhys17}; $\mathrm{UCoGe}$ under pressure is proposed as a promising candidate for $\mathbb{Z}_4$-topological superconductors~\cite{Daido_UCoGe18}.

However, M\"obius topological materials are not as well-explored as the ordinary topological materials with the local symmetry.
One of the reasons is the lack of efficient methods to evaluate topological indices for the M\"obius topological phases. For characterization of two-dimensional topological insulators without time-reversal symmetry, direct calculation of the Chern number can be done by employing the Kubo formula~\cite{TKNN,Sheng_SCH_PRL06} or the method proposed by Fukui \textit{et al.,}~\cite{Fukui_Hatsugai_05,Fukui_Hatsugai_Z2_07}, which allows us to map out phase diagrams. 
%On the other hand, so far, the method based on Wannier charge centers (WCCs)~\cite{Soluyanov_WCC_PEB2011_01,Soluyanov_WCC_PEB2011} has been employed in order to evaluate $\mathbb{Z}_4$-indices, which requires careful observations of the momentum dependence of each complicated WCC~\cite{Wang_hourglass_Nat2016,Benjamin_Wallpaperfermions_17}.  
On the other hand, the method, which has been employed to compute $\mathbb{Z}_4$-indices~\cite{Wang_hourglass_Nat2016,Benjamin_Wallpaperfermions_17}, requires carefully observing the flow of Wannier charge centers (WCCs)~\cite{Soluyanov_WCC_PEB2011_01,Soluyanov_WCC_PEB2011} for each parameter of the Hamiltonian. 
Therefore, in order to accelerate the material searching, a new technique allowing efficient computation of $\mathbb{Z}_4$-indices is strongly called for. In particular, mapping out phase diagrams is fruitful in elucidating the stability of the topological phases and also in discovering new topological states.

Under this background, we propose an efficient method directly providing $\mathbb{Z}_4$-indices both for M\"obius insulators of class AII and M\"obius superconductors of class DIII.
Specifically, with extending the method proposed by Fukui \textit{et al.}~\cite{Fukui_Hatsugai_05,Fukui_Hatsugai_Z2_07}, we introduce the $\mathbb{Z}_4$-indices defined in the lattice Brillouin zone (BZ), i.e., the descritized BZ.
Our method directly provides the $\mathbb{Z}_4$-index, for the first time. In particular, this advantage significantly improves the efficiency when one needs to obtain a phase diagram; one does not need to examine the flow of WCCs at each point of the phase diagram.
As an application, we demonstrate systematic computation of the $\mathbb{Z}_4$-index for $\mathrm{CeNiSn}$ which is proposed as a three-dimensional M\"obius Kondo insulator of class AII. 
Furthermore, we apply our method to characterize topological superconductivity for $\mathrm{UCoGe}$ whose topology has not been fully characterized. 
Our numerical analysis elucidates that the superconductivity of class DIII is strong M\"obius superconductivity for the $A_u$- or $B_{3u}$-representation.
In addition, the obtained phase diagrams 
reveal the presence of novel gapless excitations which are topologically protected by the glide symmetry; the difference of $\mathbb{Z}_4$-indices at the BZ face and at the BZ center ensures the gapless excitations as the difference of the Chern number does for the Weyl superconductor~\cite{Meng_Weylsuper_PRB12}.
We observe this type of gapless excitations with glide symmetry in a certain carrier-density region of the superconducting phase for the $B_{1u}$- or $B_{3u}$-representation.

The rest of this paper is organized as follows. In Sec.~\ref{sec: framework_main}, we show how to numerically compute the $\mathbb{Z}_4$-indices efficiently in the lattice BZ. %Brillouin zone (BZ), i.e., the discretized BZ.
In Sec.~\ref{sec: CeNiSn_main}, we apply our method to $\mathrm{CeNiSn}$ which belongs to symmetry class AII. 
We also compute the $\mathbb{Z}_4$-index by the method based on WCCs in order to show the efficiency of our method.
In Sec.~\ref{sec: UCoGe}, we apply our method to superconductivity for $\mathrm{UCoGe}$ which belongs to symmetry class DIII.
In Sec.~\ref{eq: summary}, we give a short summary.

%%%%%%%%%%%%%%%%%%%%%%%%
%\input{Framework_main.tex}
%%%%%%%%%%%%%%%%%%%%%%%%
%%%%%%%%%%%%%%%%%%%%%%%%
\section{
Framework of $\mathbb{Z}_4$-indices in the lattice BZ
}
\label{sec: framework_main}
%%%%%%%%%%%%%%%%%%%%%%%%
In Refs.~\onlinecite{Fukui_Hatsugai_05}~and~\onlinecite{Fukui_Hatsugai_Z2_07}, the lattice version of topological indices is defined in order to characterize topological insulators with local symmetry.
By extending this method, we introduce the lattice version of $\mathbb{Z}_4$-indices for class AII and class DIII.
The high efficiency of our method in the lattice BZ allows us to systematically analyze the topology of the Hamiltonian for M\"obius materials.

In the following, for each symmetry class, we introduce a lattice version of the $\mathbb{Z}_4$-index after a brief review of the $\mathbb{Z}_4$-index.
%%%%%%%%%%%%%%%%%%%%%%%%
\subsection{
Three-dimensional system of class AII
}
%%%%%%%%%%%%%%%%%%%%%%%%
%%%%%%%%%%%%%%%%%%%%%%%%
\subsubsection{
Brief review
}
%%%%%%%%%%%%%%%%%%%%%%%%
Firstly, let us review the $\mathbb{Z}_4$-index for three-dimensional systems of class AII whose BZ is cubic.
Consider a topological insulator whose Hamiltonian $H(\bm{k})$ satisfies the following constraint, 
%%%%%%%%%%%%%%%%%%
\begin{subequations}
\label{eq: AII toy model}
\begin{eqnarray}
\Theta H(\bm{k}) \Theta^{-1}&=& H(-\bm{k}), \quad \Theta^2=-1, \\
G(\bm{k}) H(\bm{k}) G(\bm{k})^{-1}&=& H(g\bm{k}), \quad G(g\bm{k})G(\bm{k})=-e^{-ik_x}, \nonumber \\
\end{eqnarray}
with
\begin{eqnarray}
\Theta G(\bm{k}) &=& G(-\bm{k})\Theta.
\end{eqnarray}
\end{subequations}
%%%%%%%%%%%%%%%%%%
Here, $\Theta$ denotes the time-reversal operator. $\bm{k}$ denotes the momentum for three-dimensional systems, $\bm{k}:=(k_x,k_y,k_z)$. $-\pi \leq k_\mu \leq  \pi$ with $\mu=x,y,z$.
The glide operator flips the momentum as $\bm{k}\to g\bm{k}:=(k_x,-k_y,k_z)$.
For $k_y=0$ or $\pi$, the Hamiltonian can be block-diagonalized with glide symmetry. 
The eigenvalues of $G(\bm{k})$ are $g_{\pm}(\bm{k})=\pm i e^{-i k_x/2}$.
Let $H_{\pm}$ denote the block-diagonalized Hamiltonian with the eigenvalue $g_{\pm}$. 

The topological structure of this phase is characterized by the $\mathbb{Z}_4$-index defined as~\cite{Shiozaki_Moebius_class_PRB16}
%%%%%%%%%%%%%%%%%%
\begin{subequations}
\begin{eqnarray}
\label{eq: Z4inv_AII}
 \theta_{3} &=& \frac{2i}{\pi} \int^\pi_{-\pi}\!dk_z \mathrm{tr}\mathcal{A}^I_{+z}(\pi,\pi,k_z) -\frac{2i}{\pi} \int^\pi_{-\pi}\!dk_z \mathrm{tr}\mathcal{A}^I_{+z}(\pi,0,k_z) \nonumber \\
             && +\frac{i}{\pi} \int^\pi_0\!dk_x \int^\pi_{-\pi}\!dk_z \mathrm{tr} \mathcal{F}_{+zx}(k_x,\pi,k_z) \nonumber \\
             && -\frac{i}{\pi} \int^\pi_0\!dk_x \int^\pi_{-\pi}\!dk_z \mathrm{tr} \mathcal{F}_{+zx}(k_x,0,k_z) \nonumber \\
             && -\frac{i}{2\pi} \int^\pi_0\!dk_y \int^\pi_{-\pi}\!dk_z \mathrm{tr} \mathcal{F}_{yz}(0,k_y,k_z) \quad (\mathrm{mod}\ 4).
\end{eqnarray}
%%%%%%%%%%%%%%%%%%
In Fig.~\ref{fig: Z4_int_domain}(a), blue colored planes indicate the region where the integrals are evaluated.
Here, $[\mathcal{A}^s_{+z}]_{nm}(\bm{k})$ and $[\mathcal{F}_{+}]_{nm}(\bm{k})$ are the Berry connection and the Berry curvature defined with the occupied states $| u^s_{+} (\bm{k}) \rangle$'s ($s=I,II$)  of the Hamiltonian $H_\pm (\bm{k})$,
%%%%%%%%%%%%%%%%%%
\begin{eqnarray}
{}[\mathcal{A}^s_{+\mu}(\bm{k})]_{nm} &=& \langle u^s_{+n}(\bm{k}) | \partial_{k_\mu} | u^s_{+m}(\bm{k}) \rangle, \label{eq: 3D_A+_cont} \\
{}[\mathcal{F}_{+\mu\nu}(\bm{k})]_{nm} &=& [\partial_{k_\mu} \mathcal{A}_{+\nu} -\partial_{k_\nu} \mathcal{A}_{+\mu}]_{nm}, \label{eq: 3D_F+_cont}
\end{eqnarray}
\end{subequations}
%%%%%%%%%%%%%%%%%%
with $\mu,\nu=x,y,z$. $s=I,II$ labels each Kramers pair.
$\mathcal{A}_{+}(\bm{k})$ denotes the Berry connection for occupied states, i.e., $\mathcal{A}_{+}(\bm{k}):=\mathcal{A}^I_{+}(\bm{k})+\mathcal{A}^{II}_{+}(\bm{k})$.
In a similar way as Eq.~(\ref{eq: 3D_F+_cont}), the Berry curvature $\mathcal{F}_{yz}$ is defined with occupied states of the Hamiltonian $H$.
We note that the $\mathbb{Z}_4$-index Eq.~(\ref{eq: Z4inv_AII}) is independent of the gauge choice modulo four.

The problem here is how to evaluate the integrals with a set of wave functions which are not defined smoothly for the BZ. 
In general, numerical simulations do not provide the smooth wave function.

%%%%%%%%%%%%%%%%%%%%%%%%
\subsubsection{
Lattice version of the $\mathbb{Z}_4$-index for symmetry class AII
}
%%%%%%%%%%%%%%%%%%%%%%%%
In this section, we show how to compute the $\mathbb{Z}_4$-index~(\ref{eq: Z4inv_AII}) with a set of wave functions which are not smooth in the BZ.
We suppose that the system satisfies Eq.~(\ref{eq: AII toy model}).

Firstly, we discretize the momentum
%%%%%%%%%%%%%%%%%%
\begin{eqnarray}
\label{eq: lattice_BZ}
(k_{x},k_{y},k_z) &=& (-\pi+i_x \Delta k,-\pi+i_y \Delta k,-\pi+i_z \Delta k), \nonumber\\
\end{eqnarray}
%%%%%%%%%%%%%%%%%%
with $i_x,i_y,i_z=0,1,\cdots,N-1$. $\Delta k=2\pi/N$. $N$ is an even integer.
Suppose that we have the block-diagonalized Hamiltonian $H_+(\bm{k})$.
Let
$
\psi(\bm{k}):=(|1(\bm{k})\rangle,|2(\bm{k})\rangle,\cdots,|2M(\bm{k})\rangle)
$
as the set of corresponding eigenvectors for occupied states.
We note that numerically obtained eigenstates are not necessarily smooth in the BZ because a gauge transformation
%%%%%%%%%%%%%%%%%%
\begin{eqnarray}
\label{eq: gauge_trans}
\psi_{n}(\bm{k})&=&\tilde{\psi}_{m}(\bm{k})V_{mn}(\bm{k}),
\end{eqnarray}
%%%%%%%%%%%%%%%%%%
is randomly applied for each point in the BZ, where $V(\bm{k})$ denotes a unitary matrix.

Even when the wave functions are not smooth in the BZ, the $\mathbb{Z}_4$-index can be computed with a proper gauge choice (see below).
In order to compute the $\mathbb{Z}_4$-index, we extend the approach proposed by Fukui \textit{et al.}~\cite{Fukui_Hatsugai_05,Fukui_Hatsugai_Z2_07}.
The lattice version of the $\mathbb{Z}_4$-index is defined as
%%%%%%%%%%%%%%%%%%
\begin{eqnarray}
\label{eq: Z4inv_FH_AII}
 \theta_{3} &=& \frac{i}{\pi} \sum_{k_z}A_{+z}(\pi,\pi,k_z)-\frac{i}{\pi} \sum_{k_z}A_{+z}(\pi,0,k_z) \nonumber \\
             && +\frac{i}{\pi} \sum_{0\leq k_x \leq \pi,k_z}F_{+zx}(k_x,\pi,k_z) \nonumber \\
             && -\frac{i}{\pi} \sum_{0\leq k_x \leq \pi,k_z}F_{+zx}(k_x,0,k_z) \nonumber \\
             && -\frac{i}{2\pi} \sum_{0\leq k_y \leq \pi,k_z}F_{yz}(0,k_y,k_z) \quad (\mathrm{mod}\ 4),
\end{eqnarray}
%%%%%%%%%%%%%%%%%%
where $\sum_{0\leq k_{x(y)} \leq \pi,k_z}$ denotes summation for $0 \leq k_{x(y)} \leq \pi$ and $-\pi \leq k_z \leq \pi$.
Here, the Berry connection and the Berry curvature are defined as
%%%%%%%%%%%%%%%%%%
\begin{subequations}
\label{eq: def_latt_A_3D}
\begin{eqnarray}
{}A_{+\mu}(\bm{k}) &=& \log U_{\mu}(\bm{k}), \\
{}F_{+\mu\nu}(\bm{k}) &=& \log U_{\mu}(\bm{k})U_{\nu}(\bm{k}+\Delta \bm{k}_\mu)U^{-1}_{\nu}(\bm{k})U^{-1}_{\mu}(\bm{k}+\Delta \bm{k}_\nu), \nonumber \\
\end{eqnarray}
%%%%%%%%%%%%%%%%%%
with 
%%%%%%%%%%%%%%%%%%
\begin{eqnarray}
U_{\mu}(\bm{k})&:=&\mathrm{det}[\psi^\dagger(\bm{k})\psi(\bm{k}+\Delta \bm{k}_\mu)]/N_\mu(\bm{k}), \\
N_\mu(\bm{k})&:=&\left|\mathrm{det}[\psi^\dagger(\bm{k})\psi(\bm{k}+\Delta \bm{k}_\mu)] \right|,
\end{eqnarray}
\end{subequations}
%%%%%%%%%%%%%%%%%%
and $\mu,\nu=x,y,z$.
$\Delta \bm{k}_x:=\Delta k(1,0,0)$. $\Delta \bm{k}_y:=\Delta k(0,1,0)$. $\Delta \bm{k}_z:=\Delta k(0,0,1)$.
We note that $A_{+\mu}$ and $F_{+\mu\nu}$ are obtained with the occupied states of $H_+$.
$F_{yz}(\bm{k})$ denotes the Berry curvature obtained with the occupied states of $H$ in a similar way as Eq.~(\ref{eq: def_latt_A_3D}b).

\begin{figure}[!h]
\begin{center}
\includegraphics[width=\hsize,clip]{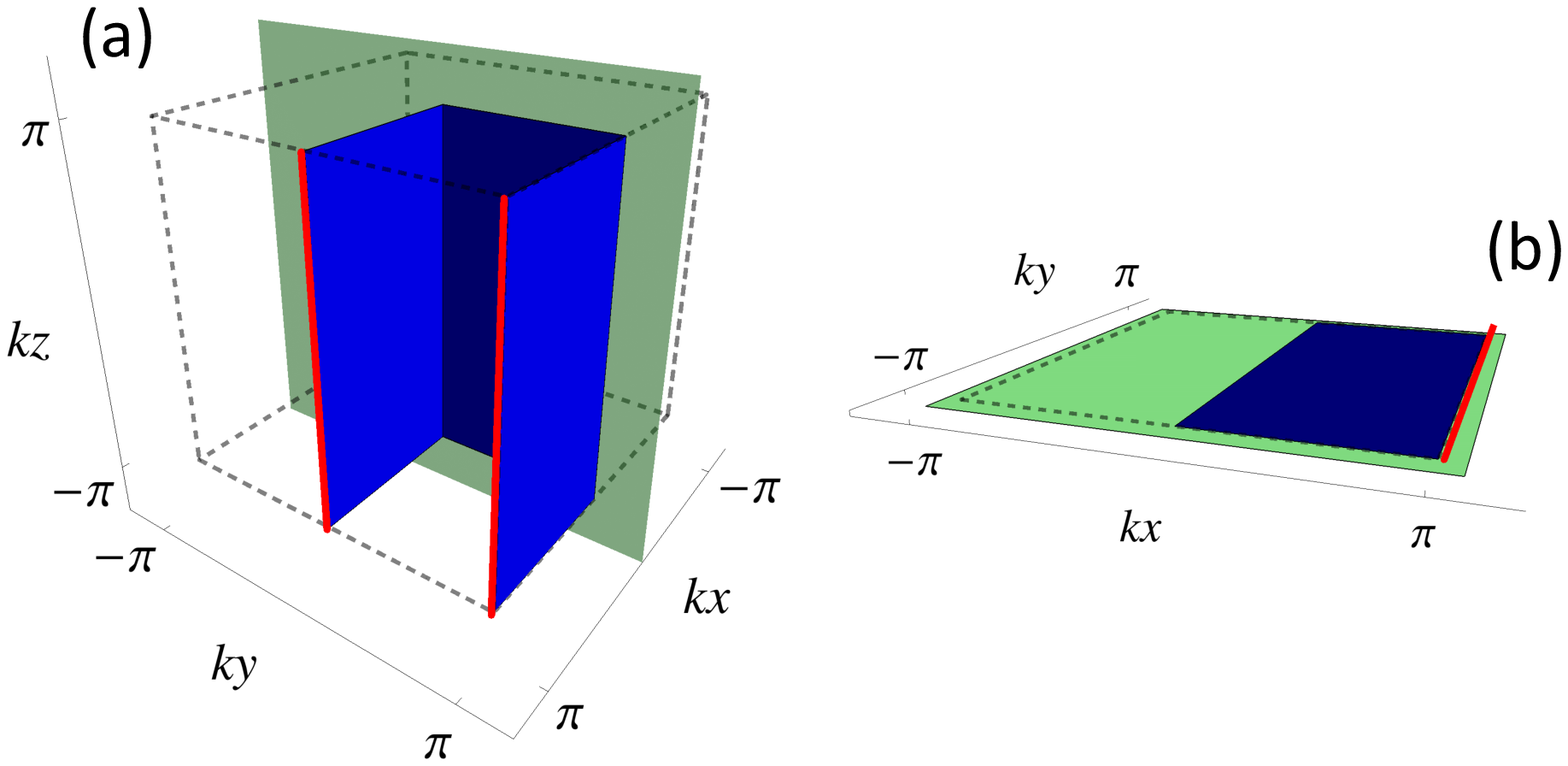}
\end{center}
\caption{(Color Online). 
(a) [(b)]: Sketch of BZ for the three- (two-) dimensional system.
Dashed lines illustrate the BZ. Blue planes denote the domain of integration. Green planes denote the glide-invariant planes. 
In the region denoted by thick red lines, the gauge is fixed as Eq.~(\ref{eq: lattice_gauge_fix_AII}) and Eq.~(\ref{eq: lattice_gauge_fix}) for three- and two- dimensional systems, respectively.
}
\label{fig: Z4_int_domain}
\end{figure}

The topological index~(\ref{eq: Z4inv_FH_AII}) is well-defined only under the following gauge fixing: along the line of $(k_x,k_y)=(\pi,0)$ or $(\pi,\pi)$ and for $-k_z<0$,
%%%%%%%%%%%%%%%%%%
\begin{subequations}
\label{eq: lattice_gauge_fix_AII}
\begin{eqnarray}
|n(-\bm{k})\rangle&=&-\Theta|n(\bm{k})\rangle,
\end{eqnarray}
and for $\bm{k}=(\pi,0,0)$, $(\pi,0,\pi)$, $(\pi,\pi,0)$, and $(\pi,\pi,\pi)$,
\begin{eqnarray}
|2n+2(\bm{k})\rangle&=&\Theta|2n+1(\bm{k})\rangle.
\end{eqnarray}
\end{subequations}
%%%%%%%%%%%%%%%%%%
In Fig.~\ref{fig: Z4_int_domain}(a), red lines denote the region where the above gauge choice should be taken.
We note that under this gauge, the integral of $\mathcal{A}^I$ in Eq.~(\ref{eq: Z4inv_AII}) can be rewritten as an integral of $\mathcal{A}$ (see Appendix~\ref{sec: AI to A}), allowing us to introduce Eq.~(\ref{eq: Z4inv_FH_AII}).

The topological index~(\ref{eq: Z4inv_FH_AII}) takes an integer modulo 4. This can be seen by noticing the following two facts.
(i)~The first and second terms take values modulo $4$ (see Appendix~\ref{sec: proof mod 4}) while the other terms are gauge independent.
(ii)~The index~(\ref{eq: Z4inv_FH_AII}) is quantized to an integer (see Appendix~\ref{sec: proof of quantization AII}).

The above two facts indicate that computing the topological index~(\ref{eq: Z4inv_FH_AII}) with the gauge choice~(\ref{eq: lattice_gauge_fix_AII}) yields the $\mathbb{Z}_4$-index of class AII.

%%%%%%%%%%%%%%%%%%%%%%%%%
%%%%%%%%%%%%%%%%%%%%%%%%%

%%%%%%%%%%%%%%%%%%%%%%%%
\subsection{
Two-dimensional systems of class DIII
}
%%%%%%%%%%%%%%%%%%%%%%%%
%%%%%%%%%%%%%%%%%%%%%%%%
\subsubsection{
Brief review
}
%%%%%%%%%%%%%%%%%%%%%%%%
Let us consider superconductivity described by the following Bogoliubov-de~Gennes (BdG) Hamiltonian $H_{\mathrm{BdG}}(\bm{k})$ satisfying,
%%%%%%%%%%%%%%%%%%
\begin{subequations}
\label{eq: symm_DIII_Z4}
\begin{eqnarray}
\Theta H_{\mathrm{BdG}}(\bm{k}) \Theta^{-1}&=& H_{\mathrm{BdG}}(-\bm{k}), \quad \Theta^2=-1, \\
C H_{\mathrm{BdG}}(\bm{k}) C^{-1}&=& -H_{\mathrm{BdG}}(-\bm{k}), \quad C^2=1, \\
G(\bm{k}) H_{\mathrm{BdG}}(\bm{k}) G(\bm{k})^{-1}&=& H_{\mathrm{BdG}}(\bm{k}), \quad G(\bm{k})^2=-e^{-ik_x}, \nonumber \\
\end{eqnarray}
with
\begin{eqnarray}
\Theta G(\bm{k}) &=& G(-\bm{k}) \Theta, \\
CG(\bm{k}) &=& -G(-\bm{k})C.
\end{eqnarray}
\end{subequations}
%%%%%%%%%%%%%%%%%%
$\Theta$ ($C$) denotes the operator of the time-reversal (particle-hole) symmetry.
The system is two-dimensional. $\bm{k}=(k_x,k_y)$. $-\pi \leq k_{x(y)} \leq  \pi$.
The glide plane is parallel to the system, and thus, the glide operator does not flip the momentum.
The eigenvalues of the matrix $G(\bm{k})$ are given by $g_{\pm}(\bm{k}):=\pm i e^{-ik_x/2}$.
The Hamiltonian can be block-diagonalized with glide symmetry. Let $H_{\pm}$ denote the block-diagonalized Hamiltonian with the eigenvalue $g_{\pm}(\bm{k}):=\pm i e^{-ik_x/2}$.

The topological structure of this phase is characterized by the $\mathbb{Z}_4$-index defined as~\cite{Shiozaki_Moebius_class_PRB16}
%%%%%%%%%%%%%%%%%%
\begin{eqnarray}
\label{eq: Z4inv}
 \theta &=& \frac{2i}{\pi} \int^\pi_{-\pi}\!dk_y \mathrm{tr}\mathcal{A}^I_{+y}(k_x=\pi,k_y), \nonumber \\
 && \quad \quad \quad \quad -\frac{i}{\pi} \int^\pi_0 \int^\pi_{-\pi}\!d^2\bm{k} \mathrm{tr} \mathcal{F}_+(\bm{k}) \quad (\mathrm{mod}\ 4).
\end{eqnarray}
%%%%%%%%%%%%%%%%%%
In Fig.~\ref{fig: Z4_int_domain}(b), the blue colored plane indicates the region where the integrals are evaluated.
The Berry connection $\mathcal{A}^I_{+}$ and the Berry curvature $\mathcal{F}_+(\bm{k})$ are defined in a similar way as Eqs.~(\ref{eq: 3D_A+_cont})~and~(\ref{eq: 3D_F+_cont}).
The $\mathbb{Z}_4$-index is well-defined for arbitrary gauge choice.

%%%%%%%%%%%%%%%%%%%%%%%%
\subsubsection{
Lattice version of the $\mathbb{Z}_4$-index for symmetry class DIII
}
%%%%%%%%%%%%%%%%%%%%%%%%
In this section, we show how to compute the $\mathbb{Z}_4$-index $\theta$ with a set of wave functions which are not smooth in the BZ.
We suppose that the system satisfies Eq.~(\ref{eq: symm_DIII_Z4}).

First, we discretize the momentum
%%%%%%%%%%%%%%%%%%
\begin{eqnarray}
\label{eq: lattice_BZ}
(k_{x},k_{y}) &=& (-\pi+i_x \Delta k,-\pi+i_y \Delta k),
\end{eqnarray}
%%%%%%%%%%%%%%%%%%
with $i_x,i_y=0,1,\cdots,N-1$, and $\Delta k=2\pi/N$. $N$ is an even integer. 
Suppose that we have the block-diagonalized Hamiltonian $H_+(\bm{k})$.
Let
$
\psi(\bm{k}):=(|1(\bm{k})\rangle,|2(\bm{k})\rangle,\cdots,|2M(\bm{k})\rangle)
$
be the set of corresponding eigenvectors for occupied states.
We again stress that numerically obtained eigenstates are not necessarily smooth in the BZ.

Even when the numerically computed states are not smooth in the BZ, the lattice version of the $\mathbb{Z}_4$-index can be computed with a proper gauge choice (see below). The lattice version of the $\mathbb{Z}_4$-index is defined as
%%%%%%%%%%%%%%%%%%
\begin{eqnarray}
\label{eq: Z4inv_FH}
 \theta &=& \frac{i}{\pi} \sum_{k_y} A_{+y}(k_x=\pi,k_y) - \frac{i}{\pi}\sum_{ 0 \leq k_x \leq \pi,k_y} F_+(\bm{k}) \quad (\mathrm{mod}\ 4). \nonumber \\
\end{eqnarray}
%%%%%%%%%%%%%%%%%%
Here, the Berry connection and the Berry curvature are defined as
%%%%%%%%%%%%%%%%%%
\begin{subequations}
\label{eq: def_latt_A}
\begin{eqnarray}
{}A_{+\mu}(\bm{k}) &=& \log U_{\mu}(\bm{k}), \\
{}F_+(\bm{k})  &=& \log U_{x}(\bm{k})U_{y}(\bm{k}+\Delta \bm{k}_x)U^{-1}_{y}(\bm{k})U^{-1}_{x}(\bm{k}+\Delta \bm{k}_y), \nonumber \\
\end{eqnarray}
%%%%%%%%%%%%%%%%%%
with 
%%%%%%%%%%%%%%%%%%
\begin{eqnarray}
U_{\mu}(\bm{k})&:=&\mathrm{det}[\psi^\dagger(\bm{k})\psi(\bm{k}+\Delta \bm{k}_\mu)]/N_\mu(\bm{k}).
\end{eqnarray}
\end{subequations}
%%%%%%%%%%%%%%%%%%
$N_\mu(\bm{k}):=|\mathrm{det}[\psi^\dagger(\bm{k})\psi(\bm{k}+\Delta \bm{k}_\mu)]|$ ($\mu=x,y$). $\Delta \bm{k}_x=\Delta k(1,0)$. $\Delta \bm{k}_y=\Delta k (0,1)$.

The topological index~(\ref{eq: Z4inv_FH}) is well-defined only under the following gauge fixing:
for $k_x=\pi$ and  $-k_y<0$, 
%%%%%%%%%%%%%%%%%%
\begin{subequations}
\label{eq: lattice_gauge_fix}
\begin{eqnarray}
|n(-\bm{k})\rangle&=&-\Theta|n(\bm{k})\rangle,
\end{eqnarray}
and for $\bm{k}=(\pi,0)$ or $(\pi,\pi)$,
\begin{eqnarray}
|2n+2(\bm{k})\rangle&=&\Theta|2n+1(\bm{k})\rangle.
\end{eqnarray}
\end{subequations}
%%%%%%%%%%%%%%%%%%
In Fig.~\ref{fig: Z4_int_domain}(b), the red line denotes the region where the above gauge choice should be taken.
Under this gauge, the integral of $\mathcal{A}^I$ can be rewritten as an integral of $\mathcal{A}$ (see Appendix~\ref{sec: AI to A}), which simplify the evaluation of the first term in Eq.~(\ref{eq: Z4inv}).

The topological index~(\ref{eq: Z4inv_FH}) takes an integer modulo 4. This can be seen by noticing the following two facts.
(i)~The first term takes a value modulo $4$ depending on the unitary matrix $V(\bm{k})$ (see Appendix~\ref{sec: proof mod 4}) while the second term is gauge independent.
(ii)~The index~(\ref{eq: Z4inv_FH}) is quantized (see Appendix~\ref{sec: proof of quantization}).

The above two facts indicate that computing the summation in Eq.~(\ref{eq: Z4inv_FH}) with the gauge choice~(\ref {eq: lattice_gauge_fix}) yields the $\mathbb{Z}_4$-index of class DIII.
We note that a $\mathbb{Z}_2$-index characterize the topology when $CG(\bm{k})=G(-\bm{k})C$ holds. The lattice version of the $\mathbb{Z}_2$-index is introduced in Appendix~\ref{sec: app_UCoGe_Z2}.

%%%%%%%%%%%%%%%%%%%%%%%%
%\input{Appl_CeNiSn.tex}
%%%%%%%%%%%%%%%%%%%%%%%%

%%%%%%%%%%%%%%%%%%%%%%%%%
\section{
Application of the $\mathbb{Z}_4$-index for class AII to $\mathrm{CeNiSn}$
}
\label{sec: CeNiSn_main}
%%%%%%%%%%%%%%%%%%%%%%%%%
Reference~\onlinecite{Chang_Moebius_CeNiSn_NatPhys17} has proposed that $\mathrm{CeNiSn}$ can be a M\"obius Kondo insulator in three dimensions. In this section, we demonstrate that the lattice version of the $\mathbb{Z}_4$-index~(\ref{eq: Z4inv_FH_AII}) efficiently characterizes the topology of the M\"obius topological Kondo insulator by the comparison with the results of the previous study.
We also obtain the $\mathbb{Z}_4$-index by computing WCCs in order to demonstrate the efficiency of our method.

%%%%%%%%%%%%%%%%%%%%%%%%%
\subsection{Model}
%%%%%%%%%%%%%%%%%%%%%%%%%
The unit cell of $\mathrm{CeNiSn}$ includes four $\mathrm{Ce}$ atoms each of which is coupled with a $\mathrm{Ni}$ atom. We label these $\mathrm{Ce}$ atoms by $a1$, $a2$, $b1$, and  $b2$. In Ref.~\onlinecite{Chang_Moebius_CeNiSn_NatPhys17}, the following effective model, of which the BZ is cubic, has been proposed to describe a M\"obius Kondo insulator with $a$-glide symmetry (see Sec.~\ref{sec: CeNiSn_symm}):
%%%%%%%%%%%%%%%%%%
\begin{subequations}
\begin{eqnarray}
\label{eq: Hami_MoebiusKondo}
\hat{H} &=& \hat{\bm{\Psi}}^\dagger_{\bm{k}} H \hat{\bm{\Psi}}_{\bm{k}}, \\
H &=& 
\left(
\begin{array}{cc}
H^c(\bm{k}) &  V(\bm{k}) \\
V^\dagger (\bm{k})  & H^f(\bm{k})
\end{array}
\right)_\lambda,
\\
\hat{\bm{\Psi}}_{\bm{k}} &=& 
\left(
\begin{array}{cc}
\hat{\bm{\Psi}}_c(\bm{k}) & \hat{\bm{\Psi}}_f(\bm{k})
\end{array}
\right),
\end{eqnarray}
%%%%%%%%%%%%%%%%%%
with 
%%%%%%%%%%%%%%%%%%
\begin{eqnarray}
H^{l(=c,f)} &=& 
\xi^l_0 +
t^l_x 
\left(
\begin{array}{cc}
0 &  \xi_x(\bm{k}) \\
\xi^\dagger_x(\bm{k}) & 0
\end{array}
\right)_{\! \tau}
\! +t^l_y 
\left(
\begin{array}{cc}
 0 & \xi_y(\bm{k}) \\
\xi^\dagger_y(\bm{k}) & 0
\end{array}
\right)_\rho,
\nonumber \\
&& \\
\xi^l_0 
&=& 
(2t^l_{z} \cos k_z  -\mu^l)s_0\tau_0\rho_0,
 \\
\xi_x(\bm{k})
&=&
s_0\left( e^{ik_x(\rho_0-\rho_z)/2} + e^{-ik_x(\rho_0+\rho_z)/2} \right),
 \\
\xi_y(\bm{k})
&=&
s_0\tau_x\left( e^{-ik_y}+1 \right),
\end{eqnarray}
%%%%%%%%%%%%%%%%%%
and 
%%%%%%%%%%%%%%%%%%
\begin{eqnarray}
V(\bm{k})&=&\!\!
\left( 
\!\!
\begin{array}{cc}
V_A(\bm{k}) & V_{AB}(\bm{k}) \\
V_{BA}(\bm{k}) & V_B(\bm{k})
\end{array}
\!\!
\right)_\rho\!\!,
\\
V_A(\bm{k}) 
&=&\!\!
\left(
\!\!
\begin{array}{cc}
2i t_2 \sin(k_z) &  t_1+s_z t_1 s_z e^{-ik_x} \\
-t_1-s_z t_1 s_z e^{ik_x}  & 2i t_2 \sin(k_z)
\end{array}
\!\!
\right)_\tau
\!\!,
\\
V_B(\bm{k}) 
&=& 
\!\!
-V_A(-\bm{k}),
\\
V_{AB} 
&=&
\!\!
\left(
\!\!
\begin{array}{cc}
0 & t_3-s_y t_3 s_y e^{-ik_y} \\
t_4-s_y t_4 s_y e^{-ik_y} & 0
\end{array}
\!\!
\right)_\tau
\!\!,
\\
V_{BA} 
&=&
\!\!
\left(
\!\!
\begin{array}{cc}
0 & s_y t_3 s_y e^{ik_y} -t_3 \\
s_y t_4 s_y e^{ik_y} -t_4 & 0
\end{array}
\!\!
\right)_\tau
\!\!,
\end{eqnarray}
\begin{eqnarray}
\hat{\bm{\Psi}}_c(\bm{k}) 
&=&
\!\!
\left(
\!\!
\begin{array}{cccc}
c_{1A\uparrow}(\bm{k}) \! & c_{1A\downarrow}(\bm{k}) \! & c_{2A\uparrow}(\bm{k}) \! & c_{2A\downarrow}(\bm{k})  
\end{array}
\right.
\nonumber \\
&&
\!\!
\!\!
\!\!
\left.
\begin{array}{cccc}
 c_{1B\uparrow}(\bm{k}) \! & c_{1B\downarrow}(\bm{k}) \! & c_{2B\uparrow}(\bm{k}) \! & c_{2B\downarrow}(\bm{k})
\end{array}
\!\!
\right)^{T}
\!\!,
\\
\hat{\bm{\Psi}}_f(\bm{k}) 
&=&
\!\!
\left(
\!\!
\begin{array}{cccc}
f_{1A\uparrow}(\bm{k}) \! & f_{1A\downarrow}(\bm{k}) \! & f_{2A\uparrow}(\bm{k}) \! & f_{2A\downarrow}(\bm{k})  
\end{array}
\right.
\nonumber \\
&&
\!\!
\!\!
\left.
\begin{array}{cccc}
 f_{1B\uparrow}(\bm{k}) \! & f_{1B\downarrow}(\bm{k}) \! & f_{2B\uparrow}(\bm{k}) \! & f_{2B\downarrow}(\bm{k}) 
\end{array}
\!\!
\right)^{T}
\!\!.
\end{eqnarray}
%%%%%%%%%%%%%%%%%%
\end{subequations}
%%%%%%%%%%%%%%%%%%
Here, the constant $t^{(l=c,f)}_{i}$ ($i=x,y,z$) denotes the hopping for each direction. $\mu^{l(=c,f)}$ represents the energy level of $c$- and $f$- electrons.
The matrix $t_i$ ($i=1,2,3,4$) is defined as
%%%%%%%%%%%%%%%%%%
\begin{subequations}
\begin{eqnarray}
 t_1 &=& i(\alpha s_x + \beta s_z), \\
 t_2 &=& i \gamma s_z, \\
 t_3 &=& i(a s_y + b s_z) , \\
 t_4 &=& i(a s_y - b s_z), 
\end{eqnarray}
\end{subequations}
%%%%%%%%%%%%%%%%%%
respectively. $a$, $b$, $\alpha$, $\beta$, and $\gamma$ are real numbers.

$\lambda_i$, $\rho_i$, $\tau_i$, and $s_i$ ($i=x,y,z$) are Pauli matrices.
$\lambda_i$'s act on the space spanned by $c$- and $f$-electrons.
$\rho_i$'s act on the space spanned by sites $1$ and $2$.
$\tau_i$'s act on the space spanned by sites $a$ and $b$.
$s_i$'s act on the spin space.

%%%%%%%%%%%%%%%%%%%%%%%%%
\subsection{
Symmetry class and crystal symmetry
}
\label{sec: CeNiSn_symm}
%%%%%%%%%%%%%%%%%%%%%%%%%
Here, we discuss the symmetry protecting the topology of the $\mathbb{Z}_4$-index.
This system preserves the time-reversal symmetry. 
The corresponding operator is given as
%%%%%%%%%%%%%%%%%%
\begin{eqnarray}
\Theta&=&is_y\lambda_0\tau_0\rho_0 \mathcal{K},
\end{eqnarray}
%%%%%%%%%%%%%%%%%%
satisfying $\Theta^2=-1$. $\mathcal{K}$ is the operator of complex conjugation.
Thus the symmetry class of this system is AII.
In addition, the Hamiltonian preserves $a$-glide symmetry $G_{xy}$ whose operator is defined as
%%%%%%%%%%%%%%%%%%
\begin{eqnarray}
\label{eq: glide_CeNiSn}
 U(G_{xy},\bm{k}) &=& 
 -i s_z
\left(
\begin{array}{cc}
                         0 & e^{ik_x(\rho_0-\rho_z)/2} \\
e^{ik_x(\rho_0+\rho_z)/2}  & 0
\end{array}
\right)_\tau,
\nonumber \\
\end{eqnarray}
%%%%%%%%%%%%%%%%%%
where $G_{xy}$ flips the momentum; $\bm{k}\to \bm{k}':=G_{xy}\bm{k}=(k_x,k_y,-k_z)$. 
We note that $ U(G_{xy},\bm{k}') U(G_{xy},\bm{k})=-e^{ik_x} $ holds. 
Thus, the Hamiltonian satisfies the condition~(\ref{eq: AII toy model}).

%%%%%%%%%%%%%%%%%%%%%%%%%
\subsection{
Numerical results
}
%%%%%%%%%%%%%%%%%%%%%%%%%
%

\begin{figure}[!h]
\begin{minipage}{1\hsize}
\begin{center}
\includegraphics[width=\hsize,clip]{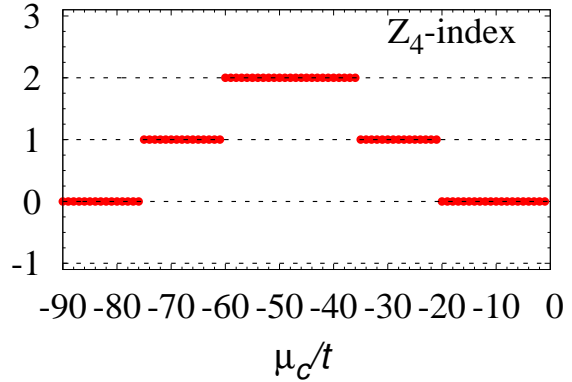}
\end{center}
\end{minipage}
\caption{(Color Online). 
$\mathbb{Z}_4$-index of the Hamiltonian~(\ref{eq: Hami_MoebiusKondo}) as a function of $\mu_c$. The data are obtained with a $120\times 120\times 120$ $k$ mesh. The data already converge for $64\times 64\times 64$ $k$ mesh. 
We note that the $\mathbb{Z}_4$-index takes $3$ for $21<\mu_c<35$.
}
\label{fig: CeNiSn}
\end{figure}

Employing the lattice version of the $\mathbb{Z}_4$-index, we characterize the topology of the system. The result is shown in Fig.~\ref{fig: CeNiSn}. 
These data are obtained for the following set of parameters,
%%%%%%%%%%%%%%%%%%
\begin{subequations}
\begin{eqnarray}
&&
\left(
\begin{array}{cccccccc}
t^c_x & t^c_y & t^c_z & \alpha &\beta & \gamma & a& b
\end{array}
\right)
\nonumber 
\\
&&\;
=
\left(
\begin{array}{cccccccc}
5 & -25 & -10 & 0.525 & 0.525 & 0.1 & 0.5 & 0.5
\end{array}
\right),
\end{eqnarray}
and 
\begin{eqnarray}
t^f_i =-t^c_i/20, &\quad& \mu_f = -\mu_c/20.
\end{eqnarray}
\end{subequations}
%%%%%%%%%%%%%%%%%%
In Fig.~\ref{fig: CeNiSn}, the computed $\mathbb{Z}_4$-index is plotted as a function of $\mu_c$.
%%%%%%%%%%%%%%%%%%%%%%%%%
%%%%%%%%%%%%%%%%%%%%%%%%%
This figure basically reproduces the one-dimensional phase diagram obtained in Ref.~\onlinecite{Chang_Moebius_CeNiSn_NatPhys17}. 
The difference can be found in the region for $-35<\mu_c<-21$. 
In this region, the previous study proposed that the $\mathbb{Z}_4$-index takes $\theta=3$ because three Dirac cones are observed at the surface of the system~\cite{Chang_Moebius_CeNiSn_NatPhys17}, while Fig.~\ref{fig: CeNiSn} indicates $\theta=1$ which is the same value as the case for $-76<\mu_c<-60$.
However, a careful observation of WCCs elucidates that the $\mathbb{Z}_4$-index takes $\theta_3=1$ in this region (for details, see Appendix~\ref{sec: WCC}), which supports our result.

Here, we discuss the efficiency of our method. As discussed in Appendix~\ref{sec: WCC}, one needs to examine the flow of WCCs to obtain the $\mathbb{Z}_4$-index for a given set of parameters. 
This means that in order to obtain a phase diagram of a realistic model, one needs to perform the procedure of the Appendix~\ref{sec: WCC} for every point in the phase diagram, which is a hard task. 
We also note that even around the transition points our data converge for $64\times64\times64$ $k$ mesh while the convergence would become worse around the transition point when the method based on WCCs is employed. 
In the above sense, we can conclude that the efficiently is significantly improved by our method.
In the next section, we systematically analyze the topological superconductivity for $\mathrm{UCoGe}$.

%%%%%%%%%%%%%%%%%%%%%%%%
%\input{Appl_UCoGe.tex}
%%%%%%%%%%%%%%%%%%%%%%%%

%%%%%%%%%%%%%%%%%%%%%%%%%
\section{
Application of the $\mathbb{Z}_4$-index for class DIII to $\mathrm{UCoGe}$
}
\label{sec: UCoGe}
%%%%%%%%%%%%%%%%%%%%%%%%%
%
%
Reference~\onlinecite{Daido_UCoGe18} has proposed that $\mathrm{UCoGe}$ under pressure is the first promising candidate of M\"obius superconductors with the $\mathbb{Z}_4$-index taking two. In that paper, a simplified representation of the $\mathbb{Z}_4$-index has been derived for the zone face of the three-dimensional BZ by using additional symmetry of the crystal. In addition, the strong and weak indices have been discussed upon a certain assumption~\cite{footnote_UCoGe}.
In spite of the above significant progress, not all the topological indices have been computed for the superconductivity in $\mathrm{UCoGe}$. In particular, in order to extract strong topological indices, it is necessary to compute the $\mathbb{Z}_4$-index at the BZ center whose simplified representation is not available due to momentum dependence of the glide operator.

In this section, by applying the lattice version of the $\mathbb{Z}_4$-index [Eq.~(\ref{eq: Z4inv_FH})], we address complete characterization of glide topological superconductivity for $\mathrm{UCoGe}$. 
Because the paring symmetry of $\mathrm{UCoGe}$ has not been identified yet, we analyze all possible pairing potentials of the parity-odd superconductivity listed in Table~\ref{table: rep}.
In order to completely characterize the topology, we compute the three-dimensional winding number $W_3$ and the $\mathbb{Z}_2$-index $\nu_\pm$ as well as the $\mathbb{Z}_4$-index $\theta$. The reason is the following. 
While the $\mathbb{Z}_4$-index of class DIII is defined on the two-dimensional BZ, $\mathrm{UCoGe}$ is a three-dimensional superconductor. 
Thus, the topology of superconducting phase is characterized by not only the $\mathbb{Z}_4$-indices but also the winding number $W_3$ [for definition, see Eq.~(\ref{eq: W_3D_app})]. 
Furthermore, although the $\mathbb{Z}_4$-index is defined when the order parameter of superconductivity has odd glide-parity (glide-odd superconductivity), superconductivity of $\mathrm{UCoGe}$ may have even glide-parity. For the glide-even superconductivity, the topology is characterized by the $\mathbb{Z}_2$-index $\nu_+$ ($\nu_-$) computed for the plus (minus) sector of glide eigenvalues [for definition, see Eq.~(\ref{eq:Z2_inv_FH_2D}) in Appendix~\ref{sec: app_UCoGe_Z2}].
Computing these topological indices elucidates whether topology of UCoGe is two- or three-dimensional one. The results are summarized in table~\ref{table: summ_s/w_UCoGe}.

To compute the topology of the superconductivity in $\mathrm{UCoGe}$, we employ the effective model proposed in Ref.~\onlinecite{Daido_UCoGe18}.

%%%%%%%%%%%%%%%%%%%%%%%%%
\subsection{Model}
\begin{figure}[!h]
\begin{center}
\includegraphics[width=\hsize,clip]{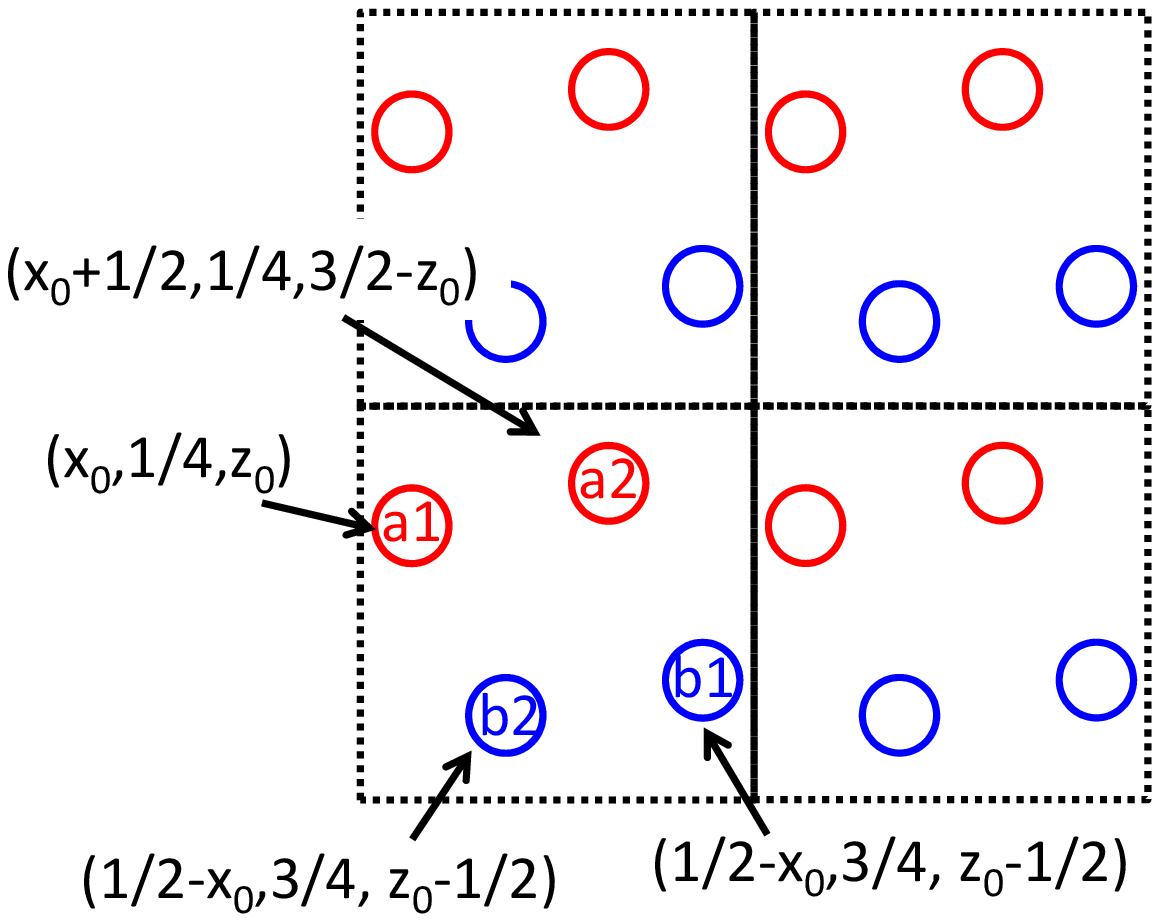}
\end{center}
\caption{(Color Online). 
Lattice structure of $\mathrm{UCoGe}$. Colored circles denote uranium atoms. Dashed black lines show the unite cell which is composed of four sites, $a1$, $a2$, $b1$, and $b2$.
Numbers with parentheses denote coordinates of atoms. Red (blue) circles are on the same plane, $y=1/4$ ($y=3/4$), respectively. $x_0=0.0101$. $z_0=0.075$~\cite{Canepa_UCoGe_crystal}.
}
\label{fig: UCoGe_latt}
\end{figure}

%%%%%%%%%%%%%%%%%%%%%%%%%
The unit cell of the system is shown in Fig.~\ref{fig: UCoGe_latt}, which is composed of four sublattices $a1$, $a2$, $b1$, and $b2$.
The effective model of the normal part is given by~\cite{Daido_UCoGe18}
%%%%%%%%%%%%%%%%%%
\begin{subequations}
\begin{eqnarray}
\hat{H}_{0}&=&
\hat{\bm{C}}^\dagger_{\bm k} [H_{\mathrm{hop}}(\bm{k})+H_{\mathrm{ASOC}}(\bm{k})] \hat{\bm{C}}_{\bm k}, 
\\
H_{\mathrm{hop}}&=& s_0
\left(
\begin{array}{cc}
H_{a} & H_{ab} \\
H^\dagger_{ab} & H^T_a
\end{array}
\right)_\eta,
\\
H_{\mathrm{ASOC}}
&=&
\bm{g}_\alpha(\bm{k})\cdot \bm{s}\sigma_z\eta_z +\bm{g}_\beta(\bm{k})\cdot \bm{s}\eta_z,\\
\hat{\bm{C}}_{\bm k}
&=&
\left(
\begin{array}{cccc}
c_{\bm{k}a1\uparrow} & 
c_{\bm{k}a1\downarrow} & 
c_{\bm{k}a2\uparrow} &
c_{\bm{k}a2\downarrow} 
\end{array}
\right.
\nonumber \\
&& \quad\quad
\left.
\begin{array}{cccc}
c_{\bm{k}b1\uparrow} &
c_{\bm{k}b1\downarrow} &
c_{\bm{k}b2\uparrow} & 
c_{\bm{k}b2\downarrow}
\end{array}
\right)^T,
\end{eqnarray}
with 
%%%%%%%%%%%%%%%%%%
\begin{eqnarray}
H_{a}
&=&
\left(
\begin{array}{cc}
\xi(\bm{k}) & \xi_{12}(\bm{k}) \\
\xi^*_{12}(\bm{k}) & \xi(\bm{k})
\end{array}
\right)_\sigma,
\\
H_{ab}&=&
\left(
\begin{array}{cc}
v_1(\bm{k}) & 0 \\
0 & v_2(\bm{k})
\end{array}
\right)_\sigma,
\\
\xi(\bm{k}) &=& 2t'_1\cos k_x +2t_2\cos k_y \nonumber \\
&& \quad +2t_3\cos k_z -\mu,\\
\xi_{12}(\bm{k}) &=& t_1(1+e^{-ik_x}),\\
v_1(\bm{k}) &=& e^{-ik_x}(1+e^{-ik_y})(t_{ab}+t'_{ab}e^{ik_z}),\\
v_2(\bm{k}) &=& e^{ik_z}(1+e^{-ik_y})(t_{ab}+t'_{ab}e^{-ik_z}),\\
\bm{g}_{\alpha} &=& 
\alpha
\left(
\begin{array}{ccc}
-\sin k_y &\delta_\alpha \sin k_x & 0
\end{array}
\right)^T, \\
\bm{g}_{\beta} &=& 
\beta
\left(
\begin{array}{ccc}
0 &\delta_{\beta} \sin k_z & \sin k_y
\end{array}
\right)^T.
\end{eqnarray}
%%%%%%%%%%%%%%%%%%
\end{subequations}
%%%%%%%%%%%%%%%%%%
Here, $s$'s, $\sigma$'s, and $\eta$'s are the Pauli matrices acting on the spin space, the sublattice space $(1,2)$, and the sublattice space $(a,b)$, respectively.
We define the matrix $H_0:=H_{\mathrm{hop}}+H_{\mathrm{ASOC}}$

Since we are interested in topological superconductivity, we consider the following BdG Hamiltonian
%%%%%%%%%%%%%%%%%%
\begin{subequations}
\begin{eqnarray}
\label{eq: UCoGe BdG}
\hat{H}_{\mathrm{super}}&=&
(\hat{\bm{C}}^\dagger_{\bm k},\hat{\bm{C}}^T_{-{\bm k}})
H_{\mathrm{BdG}}(\bm{k})
\left(
\begin{array}{c}
\hat{\bm{C}}_{\bm k}  \\
\hat{\bm{C}}^*_{- {\bm k} }
\end{array}
\right)_\tau,
\\
H_{\mathrm{BdG}}(\bm{k})
&=&
\left(
\begin{array}{cc}
H_0(\bm{k}) & \Delta(\bm{k}) \\
\Delta^\dagger(\bm{k}) & -H^T_0(-\bm{k})
\end{array}
\right)_\tau,
\end{eqnarray}
with 
\begin{eqnarray}
\hat{\bm{C}}^*_{ \bm k }
&:=&
\left(
\begin{array}{cccc}
c^\dagger_{\bm{k}a1\uparrow} & 
c^\dagger_{\bm{k}a1\downarrow} & 
c^\dagger_{\bm{k}a2\uparrow} &
c^\dagger_{\bm{k}a2\downarrow} 
\end{array}
\right.
\nonumber \\
&&\quad\quad
\left.
\begin{array}{cccc}
c^\dagger_{\bm{k}b1\uparrow} &
c^\dagger_{\bm{k}b1\downarrow} &
c^\dagger_{\bm{k}b2\uparrow} & 
c^\dagger_{\bm{k}b2\downarrow}
\end{array}
\right)^T.
\end{eqnarray}
%
%%%%%%%%%%%%%%%%%%
\end{subequations}
%%%%%%%%%%%%%%%%%%
We define $\tau$'s as the Pauli matrices acting on the Nambu space.
There are several cases of the pairing potential $\Delta(\bm{k})$ allowed by the symmetry group of $\mathrm{UCoGe}$, $Pnma$.
Under this symmetry, the four representations of odd-parity pairing are allowed. The details are summarized in Table~\ref{table: rep}~\cite{SigristUeda_91,Daido_UCoGe18}.
%%%%%%%%%%%%%%%%%%
\begin{table*}[htb]
\begin{center}
\begin{tabular}{cccccc} \hline \hline
Representation & Pairing potential $\Delta(\bm{k})$ & $a$-glide ($G_{xy}$) & $n$-glide ($G_{yz}$) & reflection ($R_y$) & inversion ($I$) \\ \hline 
$A_{u}$ & $-\Delta_x\sin k_x s_z\sigma_0\eta_0+i\Delta_y \sin k_y s_y\sigma_0\eta_0 -\Delta_z \sin k_z s_z \sigma_z\eta_0$
&odd
&odd
&odd
&odd
\\
$B_{1u}$ & $i\Delta_x\sin k_x s_0\sigma_0\eta_0-\Delta_y \sin k_y s_z\sigma_0\eta_0 +i\Delta_z \sin k_z s_0 \sigma_z\eta_0$
&odd
&even
&even
&odd
\\
$B_{2u}$ & $-\Delta_x\sin k_x s_z\sigma_z\eta_0+i\Delta_y \sin k_y s_0\sigma_z\eta_0 -\Delta_z \sin k_z s_z \sigma_0\eta_0$
&even
&even
&odd
&odd
\\
$B_{3u}$ & $i\Delta_x\sin k_x s_0 \sigma_z \eta_0 - \Delta_y \sin k_y s_z \sigma_z\eta_0 +i\Delta_z \sin k_z s_0\sigma_0\eta_0$ 
&even
&odd
&even
&odd
\\ 
\hline \hline
\end{tabular}
\caption{
Pairing potentials for each representation of point group $D_{2h}$. The third, fourth, fifth, and sixth columns indicate $\mathrm{sgn}(g)$ under the transformation of $g$(=$G_{xy}$, $G_{yz}$, $R_y$, $I$). 
For $\mathrm{sgn}(g)=1$ ($-1$) the paring potential is even (odd) under the transformation of $g$.
}
\label{table: rep}
\end{center}
\end{table*}
%%%%%%%%%%%%%%%%%%

%%%%%%%%%%%%%%%%%%%%%%%%%
%%%%%%%%%%%%%%%%%%%%%%%%%

%%%%%%%%%%%%%%%%%%%%%%%%%
\subsection{
Symmetry class and crystal symmetry
}
%%%%%%%%%%%%%%%%%%%%%%%%%
In order to see which topological indices characterize topology of the system for each paring potential in Table~\ref{table: rep}, we discuss the symmetry class and the crystal symmetry of the BdG Hamiltonian~(\ref{eq: UCoGe BdG}).

Let us start with the local symmetry.
Because the Hamiltonian preserves the time-reversal symmetry with $\Theta^2=-1$ and the particle-hole symmetry with $C^2=1$, the symmetry class is DIII.
Thus, the system can be characterized with the three-dimensional winding number $W_3$~\cite{Schnyder_classification_free_2008,Essin_Winding_PRB} whose definition is given in Eq.~(\ref{eq: W_3D_app}). 
The presence of the time-reversal symmetry and the particle-hole symmetry can be seen as follows.

\textit{(i) Time-reversal symmetry--}
The Hamiltonian~(\ref{eq: UCoGe BdG}) satisfies the time-reversal symmetry:
%%%%%%%%%%%%%%%%%%
\begin{subequations}
\label{eq: UCoGe_TRsymm}
\begin{eqnarray}
\Theta H_{\mathrm{BdG}}(\bm{k}) \Theta^{-1}&=&H_{\mathrm{BdG}}(-\bm{k}),
\end{eqnarray}
with
\begin{eqnarray}
 \Theta &=&is_y\sigma_0\eta_0\tau_0 \mathcal{K}.
\end{eqnarray}
\end{subequations}
%%%%%%%%%%%%%%%%%%

\textit{(ii) Particle-hole symmetry--}
The Hamiltonian~(\ref{eq: UCoGe BdG}) satisfies the particle-hole symmetry:
%%%%%%%%%%%%%%%%%%
\begin{subequations}
\label{eq: UCoGe_PHsymm}
\begin{eqnarray}
CH_{\mathrm{BdG}}(\bm{k})C^{-1}&=&-H_{\mathrm{BdG}}(-\bm{k}),
\end{eqnarray}
with
\begin{eqnarray}
C&=&s_0\sigma_0\eta_0\tau_x \mathcal{K}.
\end{eqnarray}
\end{subequations}
%%%%%%%%%%%%%%%%%%

As well as the local symmetry, the BdG Hamiltonian preserves the space group symmetry $Pnma$.
In order to describe the spatial symmetry, let us consider the symmetry transformation $g$ acting on the Nambu space. When the BdG Hamiltonian preserves the symmetry, it satisfies
%%%%%%%%%%%%%%%%%%
\begin{subequations}
\begin{eqnarray}
U'(g,\bm{k}) H_{\mathrm{BdG}}(\bm{k}) U'^\dagger(g,\bm{k})
&=&
H_{\mathrm{BdG}}(g\bm{k}),
\end{eqnarray}
with
\begin{eqnarray}
U'(g,\bm{k})&=&
\left(
\begin{array}{cc}
U(g,\bm{k}) & 0 \\
0 & \mathrm{sgn}(g) U^*(g,-\bm{k})
\end{array}
\right)_\tau.
\end{eqnarray}
\end{subequations}
%%%%%%%%%%%%%%%%%%
Here $U(g,\bm{k})$ is a unitary matrix satisfying
%%%%%%%%%%%%%%%%%%
\begin{subequations}
\begin{eqnarray}
U(g,\bm{k})H_0(\bm{k})U^\dagger(g,\bm{k}) &=&H_0(g\bm{k}), \\
U(g,\bm{k})\Delta(\bm{k})U^T(g,-\bm{k})&=& \mathrm{sgn}(g) \Delta(g\bm{k}),
\end{eqnarray}
\end{subequations}
%%%%%%%%%%%%%%%%%%
where $\mathrm{sgn}(g)$ takes $1$ for the even paring and takes $-1$ for the odd paring. 
In Table~\ref{table: rep}, we show $\mathrm{sgn}(g)$ for all the possible pairing states for the point group $D_{2h}$ which is associated with the space group $Pnma$. 
We have supposed that applying $g$ transforms the momentum $\bm{k}$ as $g\bm{k}$.
Now we discuss the details. 

%%%%%%%%%%%%%%%%%%
\textit{(iii) $a$-glide symmetry--}
%%%%%%%%%%%%%%%%%%
The BdG Hamiltonian preserves the $a$-glide symmetry ($g=G_{xy}$) which is described as
%%%%%%%%%%%%%%%%%%
\begin{subequations}
\begin{eqnarray}
U(G_{xy},\bm{k})
&=&(-is_z)
\left(
\begin{array}{cc}
G^a_{xy}(\bm{k}) & 0 \\
 0 & G^b_{xy}(\bm{k})
\end{array}
\right)_\eta,
\\
G^a_{xy}(\bm{k})
&=&
e^{-ik_z}
\left(
\begin{array}{cc}
0 & e^{-ik_x} \\
1 & 0
\end{array}
\right)_\sigma,
\\
G^b_{xy}(\bm{k})
&=&
\left(
\begin{array}{cc}
0         & 1 \\
e^{-ik_x} & 0
\end{array}
\right)_\sigma.
\end{eqnarray}
\end{subequations}
%%%%%%%%%%%%%%%%%%
Under the transformation, the momentum $\bm{k}$ is mapped to $G_{xy}\bm{k}:=(k_x,k_y,-k_z)$.
Thus, the system has two $a$-glide invariant planes, $k_z=0$ and $k_z=\pi$.
If the paring potential is $a$-glide-odd [$\mathrm{sgn}(G_{xy})=-1$], Eqs.~(\ref{eq: symm_DIII_Z4}d)~and~(\ref{eq: symm_DIII_Z4}e) are satisfied so that one can define $\mathbb{Z}_4$-indices, $\theta^a(0)$ and $\theta^a(\pi)$.
If the paring potential is $a$-glide-even [$\mathrm{sgn}(G_{xy})=1$], Eqs.~(\ref{eq: symm_DIII_Z2_GE}a)~and~(\ref{eq: symm_DIII_Z2_GE}b) are satisfied so that one can define $\mathbb{Z}_2$-indices, $\nu^a_\pm(0)$ and $\nu^a_\pm(\pi)$. 
Computation of $\mathbb{Z}_2$-index is explained in Appendix~\ref{sec: app_UCoGe_Z2}.

%%%%%%%%%%%%%%%%%%
\textit{(iv) $n$-glide symmetry--}
%%%%%%%%%%%%%%%%%%
The BdG Hamiltonian preserves the $n$-glide symmetry ($g=G_{yz}$) which is described as
%%%%%%%%%%%%%%%%%%
\begin{subequations}
\begin{eqnarray}
U(G_{yz},\bm{k})
&=&(-is_x)
\left(
\begin{array}{cc}
 0 & G^{ab}_{yz}(\bm{k}) \\
 e^{-ik_y-ik_z} G^{ab}_{yz}(\bm{k}')^\dagger & 0
\end{array}
\right)_\eta,
\nonumber \\
&&
\\
G^{ab}_{yz}(\bm{k})
&=&
e^{-ik_y}
\left(
\begin{array}{cc}
0         & 1 \\
e^{-ik_x} & 0
\end{array}
\right)_\sigma.
\end{eqnarray}
\end{subequations}
%%%%%%%%%%%%%%%%%%
Under the transformation, the momentum $\bm{k}$ is mapped to $\bm{k}':=(-k_x,k_y,k_z)$.
Thus, the system has two $n$-glide invariant planes, $k_x=0$ and $k_x=\pi$.
For $n$-glide-odd superconductivity [$\mathrm{sgn}(G_{yz})=-1$], one can define $\mathbb{Z}_4$-indices, $\theta^n(0)$ and $\theta^n(\pi)$.
For $n$-glide-even superconductivity [$\mathrm{sgn}(G_{yz})=1$], one can define $\mathbb{Z}_2$-indices, $\nu^n_\pm(0)$ and $\nu^n_\pm(\pi)$.

%%%%%%%%%%%%%%%%%%
\textit{(v) Reflection symmetry--}
%%%%%%%%%%%%%%%%%%
The BdG Hamiltonian preserves the reflection symmetry ($g=R_y$) which is described as
%%%%%%%%%%%%%%%%%%
\begin{eqnarray}
U(R_y,\bm{k}) &=&-is_y\sigma_0
\left(
\begin{array}{cc}
1 & 0 \\
0 & e^{-ik_y}
\end{array}
\right)_\eta.
\end{eqnarray}
%%%%%%%%%%%%%%%%%%
Under the transformation, the momentum $\bm{k}$ is mapped to $R_y\bm{k}:=(k_x,-k_y,k_z)$.
We note that for reflection-even superconductivity [$\mathrm{sgn}(R_y)=1$], the winding number in three dimensions $W_3$ is fixed to zero (see Appendix~\ref{sec: app_UCoGe_R}).

%%%%%%%%%%%%%%%%%%
\textit{(vi) Inversion symmetry--}
%%%%%%%%%%%%%%%%%%
The BdG Hamiltonian preserves the inversion symmetry ($g=I$) which is described as
%%%%%%%%%%%%%%%%%%
\begin{eqnarray}
U(I) &=&s_0 \sigma_0\eta_x.
\end{eqnarray}
%%%%%%%%%%%%%%%%%%
Under the transformation, the momentum $\bm{k}$ is mapped to $I\bm{k}:=(-k_x,-k_y,-k_z)$.
We note that for parity-odd superconductivity [$\mathrm{sgn}(I)=-1$], the value $(-1)^{W_3}$ is governed by the topology of Fermi surfaces~\cite{Sato_parityoddSC_PRB10}.

%%%%%%%%%%%%%%%%%%%%%%%%%
\subsection{Numerical results}
%%%%%%%%%%%%%%%%%%%%%%%%%
We perform the numerical calculations with setting the parameters as follows:
%%%%%%%%%%%%%%%%%%
%
%

%%%%%%%%%%%%%%%%%%
\begin{eqnarray}
&&
\left(
\begin{array}{cccccccccc}
t_1 & t_2 & t_3 & t_{ab} & t'_{ab}& t'_1& \alpha & \delta_\alpha& \beta&  \delta_\beta
\end{array}
\right)
\nonumber \\
&&=
\left(
\begin{array}{cccccccccc}
1 & 0.2 & 0.1& 0.5& 0.1& 0.1& 0.3& 0.5& 0.3& 0.5
\end{array}
\right),
\end{eqnarray}
%%%%%%%%%%%%%%%%%%
for the normal part and 
\begin{eqnarray}
\label{UCoGe_super_parameters}
\left( \Delta_x,\Delta_y,\Delta_z \right) 
&=&
\left(1,1,1 \right),
\end{eqnarray}
for the pairing potential.
%%%%%%%%%%%%%%%%%%
We note that for $\mu=0.55$ the model imitates the cylindrical Fermi surfaces of $\mathrm{UCoGe}$~\cite{Fujimori_UCoGe_PRB15,Fujimori_UCoGe_JPSJ16}.

The obtained results for $\mu=0.55$ indicate that the topological superconductivity of the $A_u$-, $B_{1u}$-, or $B_{3u}$-representation is characterized by the $\mathbb{Z}_4$-index with $\theta=2$, which is consistent with the results obtained in Ref.~\onlinecite{Daido_UCoGe18}.
Furthermore, our analysis away from $\mu=0.55$ elucidates that $\mathbb{Z}_4$-indices at the zone face and zone center predict novel gapless excitations in the bulk.
These gapless excitations are protected by difference of two topological indices, as is the case for Weyl superconductors~\cite{Meng_Weylsuper_PRB12} and line-nodal noncentrosymmetric superconductors~\cite{Yada_surface_PRB11,Sato_surface_PRB11,Schnyder_surface_PRB11,Daido_surface_PRB17}.
Interestingly, the nonsymmorphic glide symmetry plays a crucial role in the protection of the gapless excitations, in contrast to the above familiar examples protected by local symmetry.

In the following we start with an overview of the obtained results. After that we move on to the details for each case of pairing potentials.
%%%%%%%%%%%%%%%%%%%%%%%%%
\subsubsection{Overview}
%%%%%%%%%%%%%%%%%%%%%%%%%
We compute topological indices, $W_3$, $\theta$'s, and $\nu$'s for glide-odd and glide-even superconductivity.
The results for $\mu=0.55$ are summarized in Table~\ref{table: summ_index_UCoGe}. 
In this table, one can see that for the $A_u$-, $B_{1u}$-, or $B_{3u}$-representation, the system hosts M\"obius surface states characterized by the $\mathbb{Z}_4$-indices taking the value $2$.
In the presence of $a$-glide symmetry, $\mathbb{Z}_4$-indices $\theta^a(0)$ and $\theta^a(\pi)$ take 2 for the $A_u$- or $B_{1u}$-representations. 
In the presence of $n$-glide symmetry, the $\mathbb{Z}_4$-index $\theta^n(\pi)$ takes 2 for the $A_u$- or $B_{3u}$-representation.

%%%%%%%%%%%%%%%%%%
\begin{table}[htb]
\begin{center}
\begin{tabular}{ccc} \hline \hline
 & Set of topological indices & 
 topological indices 
 \\ \hline 
$A_{u}$  & $[W_{3},\theta^{a}(0),\theta^{a}(\pi),\theta^{n}(0),\theta^{n}(\pi)]$ & $[0,2,2,0,2]$ \\
$B_{1u}$ & $[W_{3},\theta^{a}(0),\theta^{a}(\pi),   \nu^{n}_+     (0),   \nu^{n}_+     (\pi)]$ & $[0,2,2,0,1]$ \\
$B_{2u}$ & $[W_{3},   \nu^{a}_+     (0),   \nu^{a}_+     (\pi),   \nu^{n}_+     (0),   \nu^{n}_+     (\pi)]$ & $[0,1,1,0,1]$ \\
$B_{3u}$ & $[W_{3},   \nu^{a}_+     (0),   \nu^{a}_+     (\pi),\theta^{n}(0),\theta^{n}(\pi)]$ & $[0,1,1,0,2]$ \\
\hline \hline 
\end{tabular}
\caption{
Topological indices for each case of pairing symmetry.
For our system, we have seen that $\mathbb{Z}_2$-indces, $\nu$'s, take the same value for the plus and the minus sector; 
$\nu^a_+(k_z)=\nu^a_-(k_z)$ for $k_z=0,\pi$ and 
$\nu^n_+(k_x)=\nu^n_-(k_x)$ for $k_x=0,\pi$.
}
\label{table: summ_index_UCoGe}
\end{center}
\end{table}
%%%%%%%%%%%%%%%%%%

%%%%%%%%%%%%%%%%%%
\begin{table}[htb]
\begin{center}
\begin{tabular}{ccc} \hline \hline 
& $a$-glide         & $n$-glide  \\ \hline \hline
$A_u$          & $(0,0,2)_o$      & $(0,1,0)_o$  \\
$B_{1u}$       & $(0,0,2)_o$       &  $(0,1,0)_e$ \\
$B_{2u}$       & $(0,0,1)_e$ &  $(0,1,0)_e$ \\
$B_{3u}$       & $(0,0,1)_e$ &  $(0,1,0)_o$ \\ \hline \hline
\end{tabular}
\caption{
Strong and weak topological indices.
$(\mathbb{Z},\mathbb{Z}^{\mathrm{strong}}_2,\mathbb{Z}^{\mathrm{weak}}_4)_o$ for glide-odd superconductivity and 
$(\mathbb{Z}^{\mathrm{weak}}_2,\mathbb{Z}^{\mathrm{strong}}_2,\mathbb{Z}^{\mathrm{weak}}_2)_e$ for glide-even superconductivity.
Generators of each group are summarized in Table~\ref{table: summ_UCoGe_gen}.
The data are obtained for $\mu=0.55$ where the normal part of the effective model reproduces the Fermi surface of $\mathrm{UCoGe}$.
}
\label{table: summ_s/w_UCoGe}
\end{center}
\end{table}
%%%%%%%%%%%%%%%%%%

%
Based on the above results, we can extract strong and weak topological indices with glide symmetry. The results are summarized in Table~\ref{table: summ_s/w_UCoGe} which shows nice agreement with the results obtained in Ref.~\onlinecite{Daido_UCoGe18}.
Firstly, we discuss a generic topological classification of glide symmetric superconductivity in three dimensions.
In Ref.~\onlinecite{Yanase_Moebius_UPt3_PRB17}, it has been pointed out that topological superconductivity for given glide symmetry forms an Abelian group; $(\mathbb{Z},\mathbb{Z}^{strong}_2,\mathbb{Z}^{weak}_4)_o$ for glide-odd superconductivity and $(\mathbb{Z}^{weak}_2,\mathbb{Z}^{strong}_2,\mathbb{Z}^{weak}_2)_e$ for glide-even superconductivity. 
The generators of these groups are phases described by $\mathcal{H}_{1,o(e)}$, $\mathcal{H}_{2,o(e)}$, and $\mathcal{H}_{3,o(e)}$ for the glide-odd (glide-even) superconductivity, respectively.
Here $\mathcal{H}_{i,o}$ $(i=1,2,3)$ [$\mathcal{H}_{i,e}$ $(i=1,2,3)$] describes fully gapped superconducting phases with topological indices listed in the top (bottom) panel of Table~\ref{table: summ_UCoGe_gen}.

We note that topology of superconductivity described by $\mathcal{H}_{i,o(e)}$ $(i=2,3)$ is protected by glide symmetry. 
In particular, topological surface states of $\mathcal{H}_{2,o}$ or $\mathcal{H}_{2,e}$ are protected by the strong index while those of $\mathcal{H}_{3,o}$ or $\mathcal{H}_{3,e}$ are protected by the weak index.
Therefore, surface states of topological superconductivity $\mathcal{H}_{2,o}$ or $\mathcal{H}_{2,e}$ are robust against disorder if glide symmetry is preserved on average~\cite{Mong_disorder_PRL12,Ringel_disorder_PRB12,Fu_disorder_PRB12,Fluga_disorder_PRB14} while those of $\mathcal{H}_{3,e}$ are fragile. 
We note that the stability of weak topological crystalline superconductivity is different between $\mathbb{Z}_2$- and $\mathbb{Z}_4$-classifications.
Namely, in the case of the glide-odd superconductivity $\mathcal{H}_{3,o}$, the surface states survive because hybridizing two subsystems of topological superconductivity with $\theta=1$, which arise from the BZ face and the BZ center, yields the superconductivity with $\theta=2$.
%
%%%%%%%%%%%%%%%%%%
\begin{table}[htb]
\begin{center}
\begin{tabular}{cccccc} \hline \hline
                 & $W_{3}$ & $\theta(0)$  & $\theta(\pi)$ &  &group structure \\ \hline \hline
$\mathcal{H}_{1,o}$  &   1     &   1            &    0        &  & $\mathbb{Z}$             \\
$\mathcal{H}_{2,o}$  &   0     &   0            &    2        &  & $\mathbb{Z}^{\mathrm{strong}}_2$           \\
$\mathcal{H}_{3,o}$  &   0     &   1            &    1        &  &$\mathbb{Z}^{weak}_4$    \\ \hline \hline
                 & $\nu_+(0)$ & $\nu_-(0)$  & $\nu_+(\pi)$  & $\nu_-(\pi)$  & group structure \\ \hline \hline
$\mathcal{H}_{1,e}$  &   1     &   0        &   1           &    0          & $\mathbb{Z}^{\mathrm{weak}}_2$ \\
$\mathcal{H}_{2,e}$  &   0     &   0        &   1           &    1          & $\mathbb{Z}^{\mathrm{strong}}_2$ \\
$\mathcal{H}_{3,e}$  &   1     &   1        &   1           &    1          & $\mathbb{Z}^{\mathrm{weak}}_2$  \\ \hline \hline
\end{tabular}
\caption{
(Top panel): Generators of Abelian groups formed by glide-odd superconductivity. $\mathbb{Z}_4$-indices $\theta(0)$ and $\theta(\pi)$ are defined for $a$-glide or $n$-glide symmetry.
(Bottom panel): Generators of Abelian groups formed by glide-even superconductivity.
$\mathbb{Z}_2$-indices $\nu$'s are defined for $a$-glide or $n$-glide symmetry.
The topological phase generated by $\mathcal{H}_{1,o}$, $\mathcal{H}_{2,o}$, or $\mathcal{H}_{2,e}$ cannot be obtained by stacking two-dimensional topological superconductors in the momentum space, while the one generated by $\mathcal{H}_{3,o}$, $\mathcal{H}_{1,e}$, or $\mathcal{H}_{3,e}$ can be obtained by stacking two-dimensional topological superconductors.
}
\label{table: summ_UCoGe_gen}
\end{center}
\end{table}
%%%%%%%%%%%%%%%%%%
%

Following the above generic argument, we discuss the strong topological superconductivity for $\mathrm{UCoGe}$ (see Table~\ref{table: summ_s/w_UCoGe}).
Concerning $n$-glide-odd superconductivity, we can see that the topological superconductivity of the $A_u$- or $B_{3u}$-representation is characterized by $(0,1,0)_o$ in Table~\ref{table: summ_s/w_UCoGe}, indicating that this superconductivity is topologically identical to the strong topological superconductivity described by $\mathcal{H}_{2,o}$.
Thus, the M\"obius surface states for the $A_u$- or $B_{3u}$-representation are protected by the strong index and are robust against disorder (for more details see Sec.~\ref{sec: UCoGe_B3u_n-glide}).
Concerning $n$-glide-even superconductivity, Table~\ref{table: summ_s/w_UCoGe} shows that surface states for the $B_{1u}$- or $B_{2u}$-representation are protected by the strong index and are robust against disorder.
Table~\ref{table: summ_s/w_UCoGe} indicates that $a$-glide symmetry of $\mathrm{UCoGe}$ protects only topological properties characterized by the weak index.

%%%%%%%%%%%%%%%%%%%%%%%%%
\begin{figure}[!h]
\begin{minipage}{1.0\hsize}
\begin{center}
\includegraphics[width=90mm,clip]{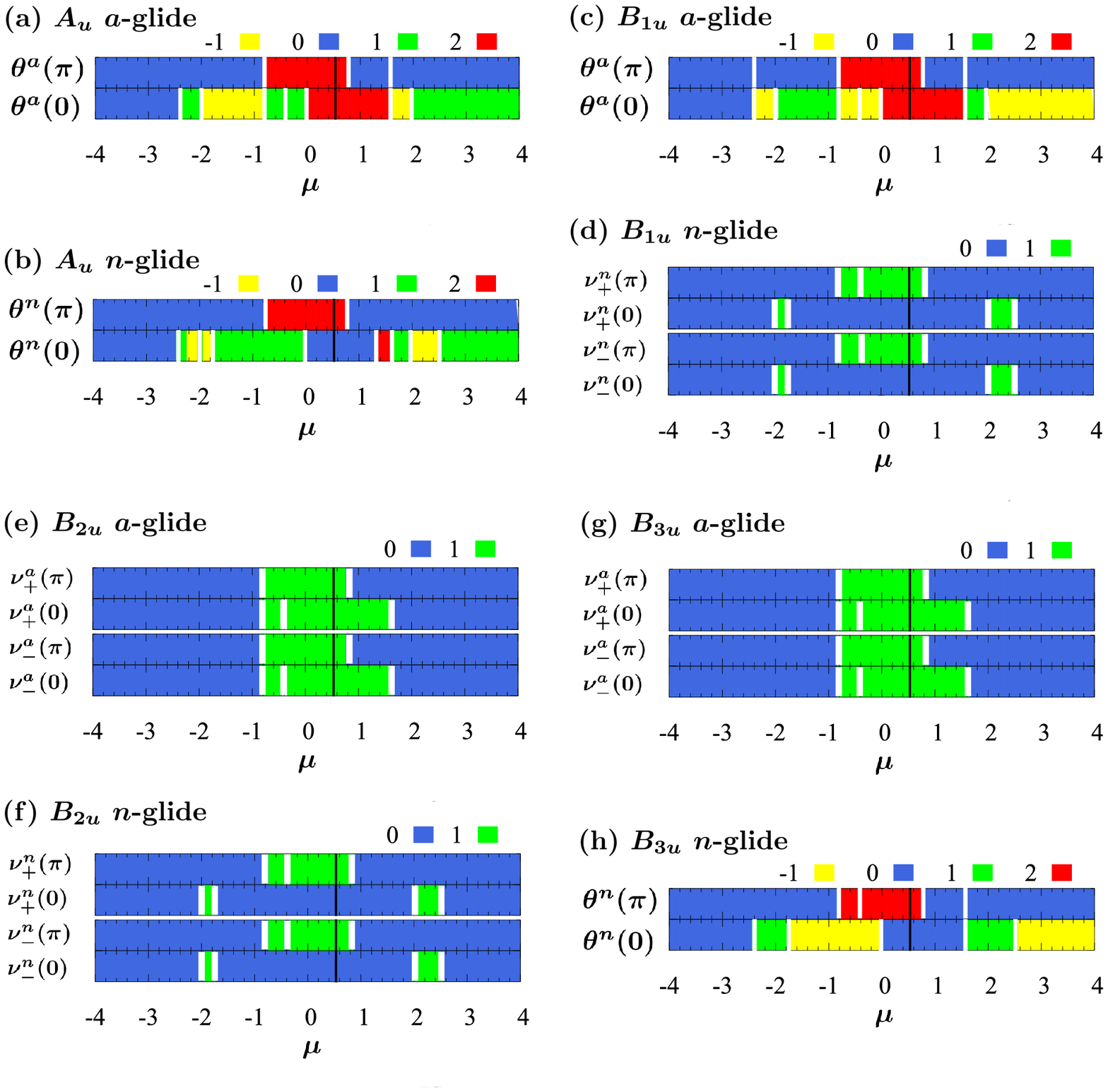}
\end{center}
\end{minipage}
\begin{minipage}{0.5\hsize}
\begin{center}
\includegraphics[width=\hsize,clip]{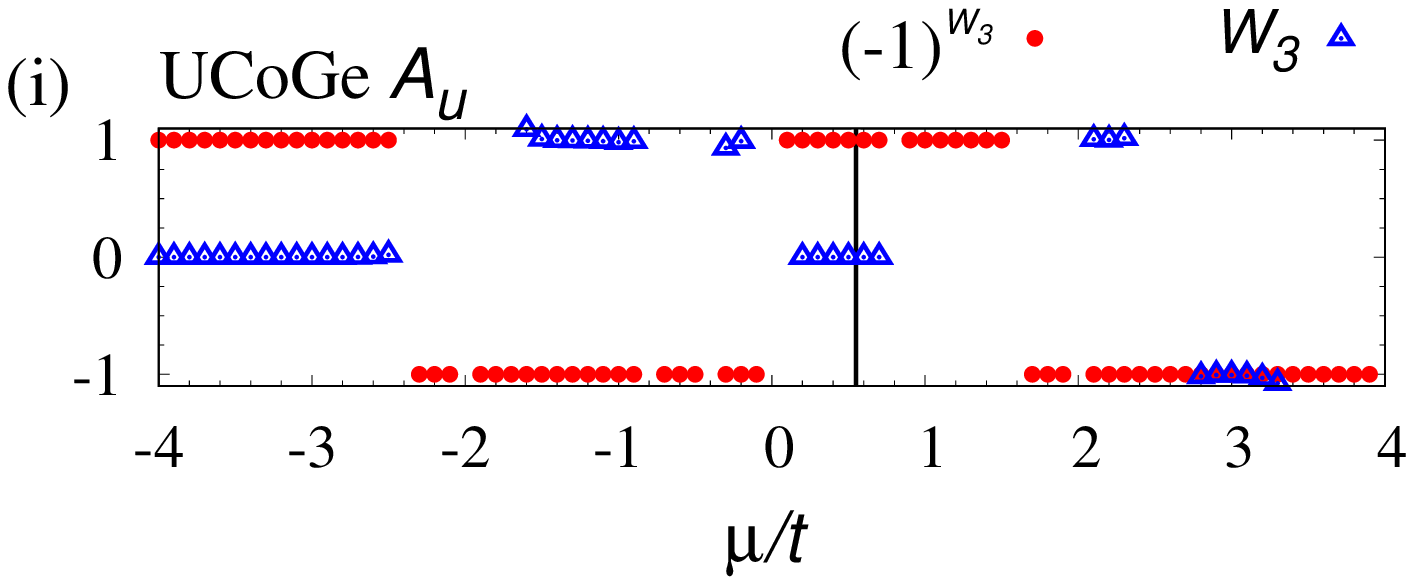}
\end{center}
\end{minipage}
\caption{(Color Online). 
$\mathbb{Z}_4$- and $\mathbb{Z}_2$-indices as functions of the chemical potential 
for the $A_u$-representation [(a) and (b)],
for the $B_{1u}$-representation [(c) and (d)],
for the $B_{2u}$-representation [(e) and (f)],
and 
for the $B_{3u}$-representation [(g) and (h)].
The black lines in these figures denote the point $\mu=0.55$ where the effective model mimics the cylindrical Fermi surfaces of $\mathrm{UCoGe}$. 
The results at this parameter are listed in Table~\ref{table: summ_index_UCoGe}.
The white lines denote the points where the indices are ill-defined because of gap-closing in the domain of integration in the BZ.
The data are obtained with a $120\times120$ $k$ mesh.
The panel (i) represents the winding number $W_3$ and its parity $(-1)^{W_3}$. 
In the following regions, $W_3$ could not be obtained with sufficient accuracy because the bulk gap is narrow:
$-2.5<\mu<-1.7$, $-0.8<\mu<-0.4$, $-0.1<\mu<0.1$, $0.7<\mu<2.2$, and $3.8<\mu \leq 4$.
We note that the parity $(-1)^{W_3}$ can be calculated only from data of normal states for parity-odd superconductivity~\cite{Sato_parityoddSC_PRB10}.
The winding number is zero ($W_3=0$) or is ill-defined for $B_{1u}$-, $B_{2u}$-, and $B_{3u}$- representations.
}
\label{fig: UCoGe_phase}
\end{figure}
%%%%%%%%%%%%%%%%%%%%%%%%%

Now, let us briefly discuss the case away from $\mu=0.55$, 
where the $\mathbb{Z}_4$-indices protect novel gapless excitations in the bulk.
The top panel of Table~\ref{table: summ_UCoGe_gen} indicates that topological indices of glide-odd superconductivity must satisfy the following condition when the system is fully gapped:~$\theta(0)+ \theta(\pi)= W_3$~(mod~$2$).
In other words, topologically protected gapless excitations appear in the bulk when the parities of two values, $\theta(0)+\theta(\pi)$ and $W_3$ are incompatible with each other.
Topological indices as functions of $\mu$ for each case of paring symmetry are summarized in Fig.~\ref{fig: UCoGe_phase}.
This result shows that for the $B_{1u}$- or $B_{3u}$-representation, doping holes (decreasing chemical potential) results in gapless excitations in the bulk predicted by the $\mathbb{Z}_4$-indices (see Sec.~\ref{sec: UCoGe_B3u_n-glide}).
The above gapless excitations with glide symmetry can be regarded as an extension of the excitations of Weyl superconductors in the following sense.
In both cases, gapless excitations are protected by difference of two topological indices. 
Namely, the $\mathbb{Z}_4$-indices protect the gapless excitations in our case, while the $\mathbb{Z}$-indices protect the gapless excitations~\cite{footnote_Z4Weyl} in Weyl~\cite{Meng_Weylsuper_PRB12} and line-nodal superconductors~\cite{Yada_surface_PRB11,Sato_surface_PRB11,Schnyder_surface_PRB11,Daido_surface_PRB17}. 
Furthermore, the glide symmetry plays a crucial role in protection of gapless excitations for the former case in contrast to those for the latter case~\cite{Meng_Weylsuper_PRB12,Yada_surface_PRB11,Sato_surface_PRB11,Schnyder_surface_PRB11,Daido_surface_PRB17}.

In the following, we discuss the above results in details. Firstly, we discuss topology of superconductivity with the $B_{3u}$-representation. 
After that we address the other cases of paring symmetry.

%%%%%%%%%%%%%%%%%%
\subsubsection{
Topology for the $B_{3u}$-representation
}
\label{sec: UCoGe_B3u_n-glide}
%%%%%%%%%%%%%%%%%%
The paring potential of $B_{3u}$-representation is even (odd) under the $a$- ($n$-) glide transformation, respectively. Thus, we discuss each case separately.

%%%
(i) \textit{$n$-glide symmetry--}
%%%
As seen in Table~\ref{table: rep}, the pairing potential is odd under the $n$-glide transformation, meaning that the topology protected by $n$-glide symmetry is completely characterized by $W_3$, $\theta^n(0)$, and $\theta^n(\pi)$~\cite{Daido_UCoGe18,Yanase_Moebius_UPt3_PRB17}.
The pairing potential of the $B_{3u}$-representation is reflection-even, which fixes the winding number to zero (for details see Appendix~\ref{sec: app_UCoGe_R}).

\begin{figure}[!h]
\begin{minipage}{0.45\hsize}
\begin{center}
\includegraphics[width=\hsize,clip]{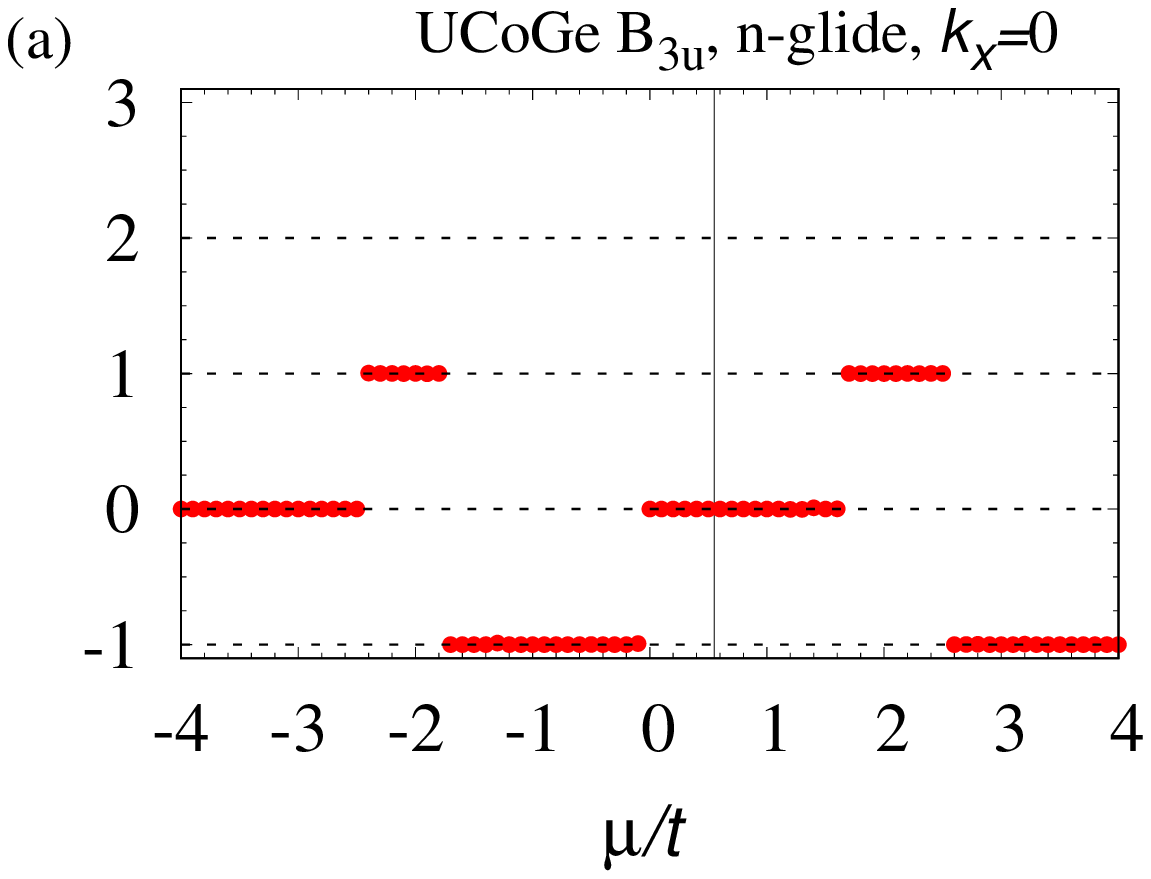}
\end{center}
\end{minipage}
\begin{minipage}{0.45\hsize}
\begin{center}
\includegraphics[width=\hsize,clip]{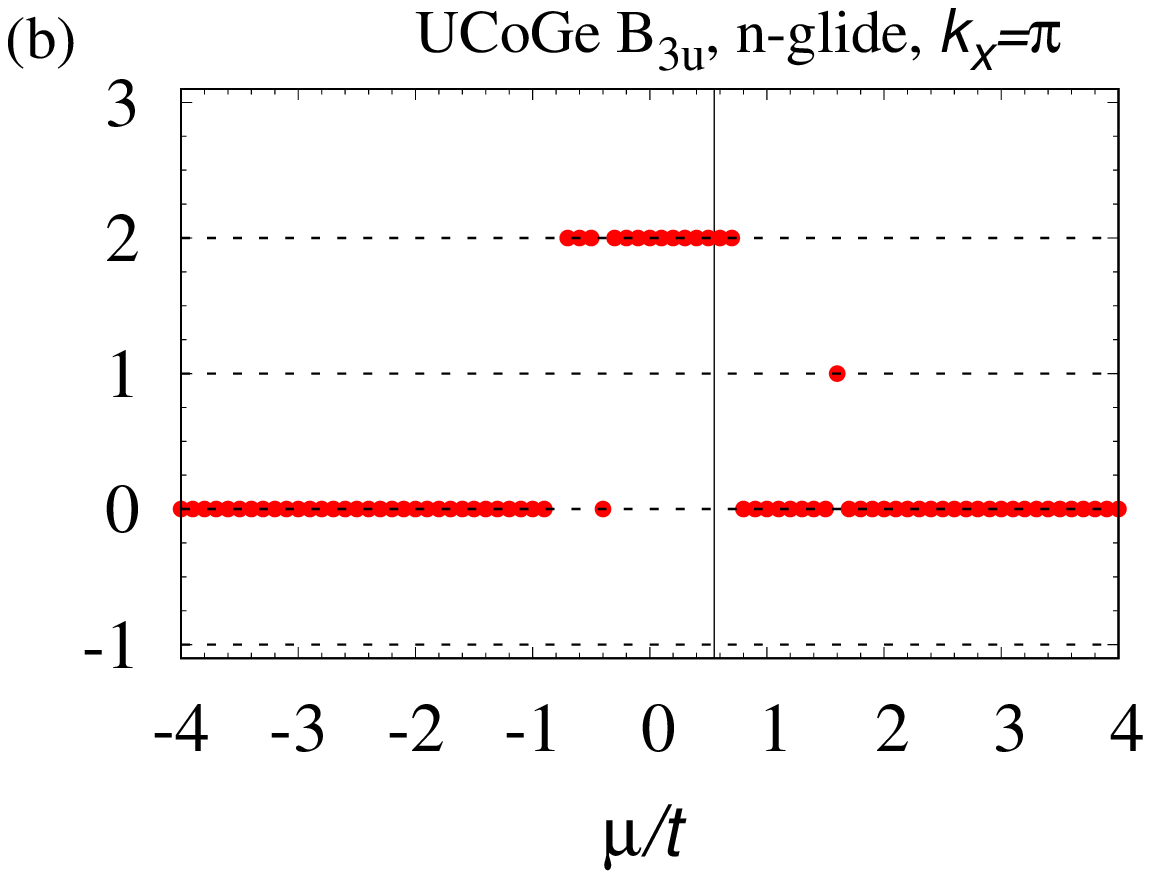}
\end{center}
\end{minipage}
\caption{(Color Online). 
$\mathbb{Z}_4$-index for the pairing potentials of the $B_{3u}$-representation.
(left panel): $\mathbb{Z}_4$-index at $k_x=0$.
(right panel): $\mathbb{Z}_4$-index at $k_x=\pi$.
At $k_x=\pi$, the gap closes for $\mu=-0.8,-0.4,0.8,1.6$.
At $k_x=0$, the gap closes for $\mu=-2.4,(-1.8\sim-1.7),0,1.6,(2.5\sim2.6)$.
Isolated dots in the right panel are due to gap-closing in the bulk.
}
\label{fig: B3umu_indx}
\end{figure}

Now, we discuss the computed $\mathbb{Z}_4$-indices which are plotted as functions of chemical potential $\mu$ in Fig.~\ref{fig: B3umu_indx}.
%%%%%%%%%%%%%%%%%%%%%%%%%
%%%%%%%%%%%%%%%%%%%%%%%%%
The black vertical lines denote the point $\mu=0.55$ where the normal part mimics the cylindrical Fermi surfaces of $\mathrm{UCoGe}$.
The figure shows that at this point, the $\mathbb{Z}_4$-indices take $[\theta^n(0),\theta^n(\pi)]=[0,2]$.
By taking into account the vanishing winding number ($W_{3}=0$), we see that the topology in this case is characterized with $[W_{3},\theta^n(0),\theta^n(\pi)]=[0,0,2]$, meaning that this topological phase is identical to the phase described by $n\mathcal{H}_{1,o}\oplus m \mathcal{H}_{2,o}\oplus l \mathcal{H}_{3,o}$ with $(n,m,l)_{o}=(0,1,0)_{o}$. 
Here, $n\mathcal{H}_{1,o}$ denotes $n$-copies of a topological phase described by the Hamiltonian $\mathcal{H}_{1,o}$. 
The direct sum denotes stacking topological phases and introducing small perturbations which do not change the topology. 
Because topology of the system is characterized by the indices $(0,1,0)_{o}$, the system hosts the M\"obuis surface states only for $k_x=\pi$, preventing the hybridization with the surface states at $k_x=0$. Thus, for the $B_{3u}$-representation, the $n$-glide-odd superconductivity is protected by the strong topological index. 
Namely, the strong M\"obius topological superconductivity is realized.

%%%%%%%%%%%%%%%%%%%%%%%%%
\begin{figure}[!h]
\begin{minipage}{0.45\hsize}
\begin{center}
\includegraphics[width=\hsize,clip]{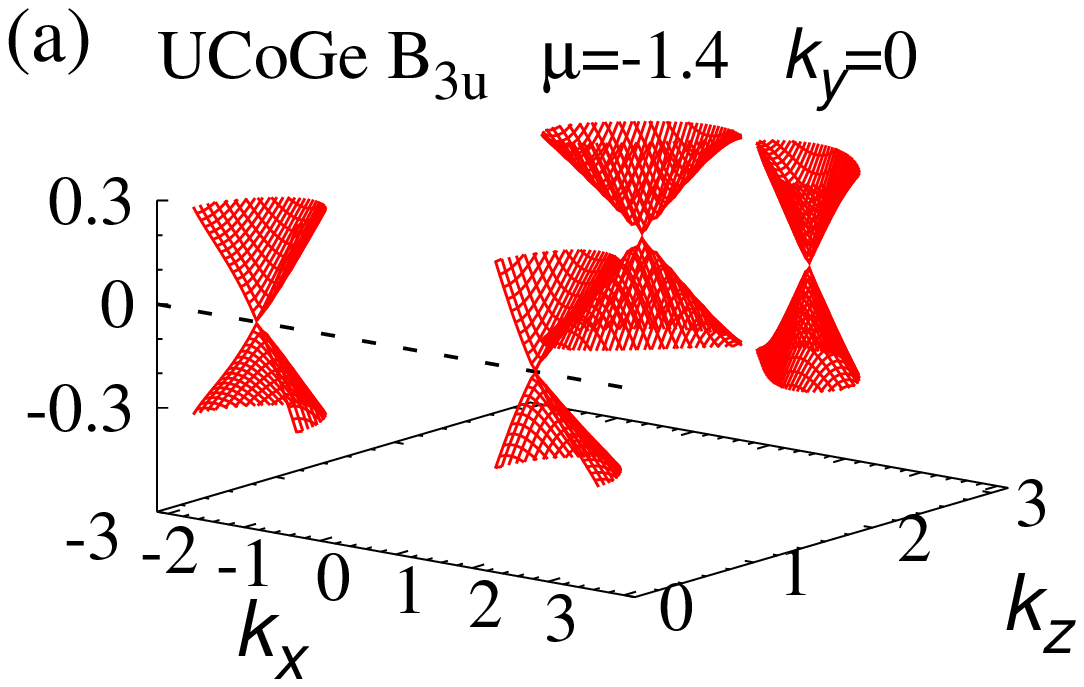}
\end{center}
\end{minipage}
\begin{minipage}{0.45\hsize}
\begin{center}
\includegraphics[width=\hsize,clip]{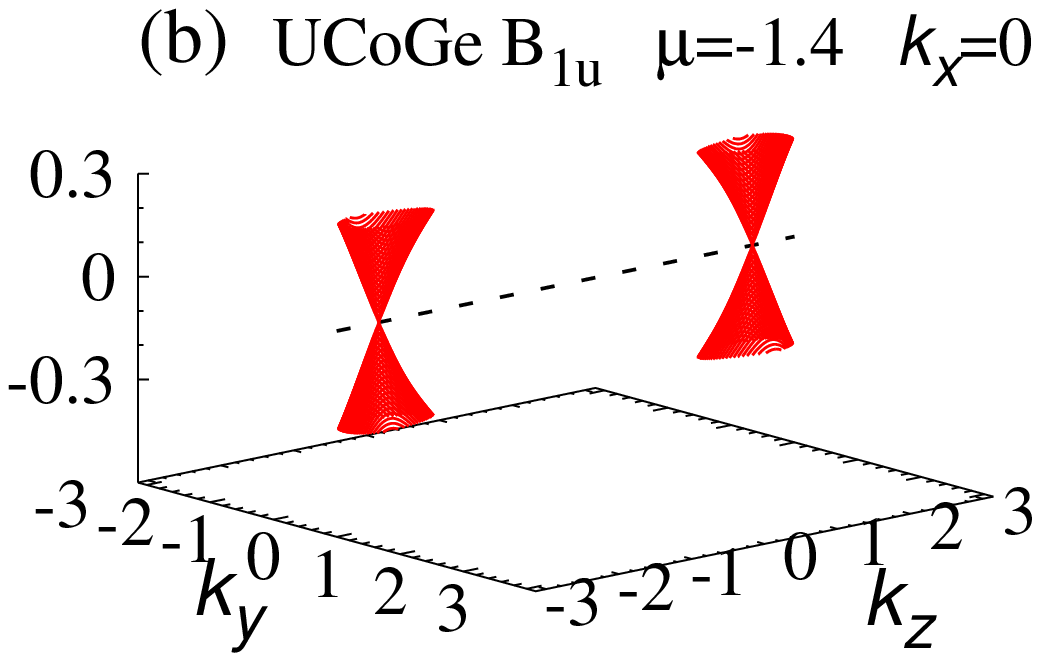}
\end{center}
\end{minipage}
\caption{(Color Online). 
(a): Bulk energy spectrum of the BdG Hamiltonian in the low energy region for the $B_{3u}$-representation at $k_y=0$.
(b): Bulk energy spectrum of the BdG Hamiltonian in the low energy region for the $B_{1u}$-representation at $k_x=0$.
The dashed lines denote high symmetry lines where the pairing potential is forbidden.
}
\label{fig: Weyl_B3u_B1u}
\end{figure}
%%%%%%%%%%%%%%%%%%%%%%%%%

Here, our systematic computation reveals that the difference of the $\mathbb{Z}_4$-indices requires the gapless excitations in the bulk as the difference of the Chern number does for Weyl superconductivity.
In Fig.~\ref{fig: B3umu_indx}, we can see that doping holes (decreasing the chemical potential) yields the phase with $[\theta^n(0),\theta^n(\pi)]=[-1,2]$ for $-0.8<\mu<0$ which cannot be generated from fully gapped superconductivity $\mathcal{H}_{1,o}$, $\mathcal{H}_{2,o}$, and $\mathcal{H}_{3,o}$ due to incompatibility between the parity of $\theta^n(0)+ \theta^n(\pi)$ and that of the winding number $W_3$.
Therefore in this region, the $\mathbb{Z}_4$-indices protected by $n$-glide symmetry require gapless excitations in the bulk BZ sandwiched by planes at $k_z=0$ and $\pi$.
Interestingly, the nonsymmorphic glide symmetry plays an essential role in the protection of these gapless excitations in contrast to well-known examples, such as Weyl superconductivity and line-nodal superconductivity where crystal symmetry is not needed.
The energy spectrum for the bulk [Fig.~\ref{fig: Weyl_B3u_B1u}(a)] support this scenario where one can observe gapless point nodes.
Classification of pairing potentials provides another insight into the gapless point-nodal excitations protected with the glide symmetry. 
Suppose that the paring potential is sufficiently small compared to the energy difference of the bands for normal states. Then, the gapless excitations appear at points on the Fermi surface where the paring potential is forbidden by symmetry.
Our analysis summarized in Appendix~\ref{sec: app_UCoGe_class_pairing_lines} elucidates that paring potential of the $B_{3u}$-representation is forbidden along the line for $(k_y,k_z)=(0,0)$ illustrated by a dashed line in Fig.~\ref{fig: Weyl_B3u_B1u}(a). 
We note that similar gapless excitations in the bulk can be observed for $-2.4<\mu<-0.8$ and for $1.6<\mu<2.5$.

%%%
(ii) \textit{$a$-glide symmetry--}
%%%
As seen in Table~\ref{table: rep}, the system can show $a$-glide-even superconductivity characterized by $\nu^a_\pm(0)$ and $\nu^a_\pm(\pi)$.
Fig.~\ref{fig: UCoGe_phase}(g) indicates that for $\mu=0.55$ the topological indices take $\nu^a_\pm(0)=\nu^a_\pm(\pi)=1$, meaning that the system is characterized by $(0,0,1)_{e}$.
Therefore, for the $B_{3u}$-representation, the system shows weak topological crystalline superconductivity protected by $a$-glide symmetry.

%%%%%%%%%%%%%%%%%%
\subsubsection{Topology for the $A_{u}$-representation}
%%%%%%%%%%%%%%%%%%
As seen in Table~\ref{table: rep}, the paring potential is odd under the $a$- and $n$-glide transformations.
Therefore, the topology is characterized by the winding number and $\mathbb{Z}_4$-indices; $W_3$, $\theta^{a(n)}(0)$, and $\theta^{a(n)}(\pi)$ for $a$- ($n$-) glide symmetry, respectively~\cite{Daido_UCoGe18,Yanase_Moebius_UPt3_PRB17}.

%%%
(i) \textit{$a$-glide symmetry--}
%%%
Fig.~\ref{fig: UCoGe_phase}(a) shows the $\mu$ dependence of $\mathbb{Z}_4$-indices. 
This figure indicates that for $\mu=0.55$, the indices take $\theta^a(0)=\theta^a(\pi)=2$. Besides, our direct calculation shows $W_3=0$ at this parameter [Fig.~\ref{fig: UCoGe_phase}(i)]. 
These two facts indicate that the system is characterized by  $[W,\theta^a(0),\theta^a(\pi)]=[0,2,2]$ which can be represented as $(0,0,2)_{o}$. 
Therefore, for the $A_u$-representation, the system shows weak topological crystalline superconductivity protected by $a$-glide symmetry.

Now we discuss the case away from $\mu=0.55$.
We note that the parity of the winding number is governed by the Fermi surface~\cite{Sato_parityoddSC_PRB10} as the system of the $A_u$-representation is parity-odd superconductivity. 
The computed parity, $(-1)^{W_3}$, is plotted as a function of $\mu$ in Fig.~\ref{fig: UCoGe_phase}(i). 
In contrast to the glide-odd superconductivity with the $B_{3u}$-representation, the parity of $\theta^a(0)+\theta^a(\pi)$ is compatible with that of the winding number for $-4<\mu<4$. 
Therefore, changing chemical potential does not yield the topological gapless phases protected with the glide symmetry.

%%%
(ii) \textit{$n$-glide symmetry--}
%%%
In a similar way to the previous case, we discuss the topology protected by $n$-glide symmetry, which is characterized by $W_3$, $\theta^{n}(0)$, and $\theta^{n}(\pi)$ as mentioned above.

Fig.~\ref{fig: UCoGe_phase}(b) indicates that for $\mu=0.55$, the system is characterized by $[W,\theta^n(0),\theta^n(\pi)]=[0,0,2]$ which can be represented as $(0,1,0)_{o}$.
Therefore, for the $A_u$-representation, the system shows strong topological crystalline superconductivity protected by $n$-glide symmetry.

As in the case of $a$-glide symmetry, changing chemical potential does not yield the topological gapless phases.

%%%%%%%%%%%%%%%%%%
\subsubsection{Topology for the $B_{1u}$-representation}
\label{sec: UCoGe_B1u}
%%%%%%%%%%%%%%%%%%
%%%
(i) \textit{$a$-glide symmetry--}
%%%
As seen in Table~\ref{table: rep}, the paring potential is odd under the $a$-glide transformation. 
Thus, topology protected by $a$-glide symmetry is characterized by the winding number and $\mathbb{Z}_4$-indices, $W_3$, $\theta^a(0)$, and $\theta^a(\pi)$.
We note that the system shows reflection-even superconductivity, fixing the winding number to zero (see Appendix~\ref{eq: W_3D_app_ref}).

Figure~\ref{fig: UCoGe_phase}(c) indicates that $\mathbb{Z}_4$-indices take $\theta^a(0)=\theta^a(\pi)=2$ for $\mu=0.55$.
Then, for $\mu=0.55$ the superconductivity is characterized by $[W,\theta^a(0),\theta^a(\pi)]=[0,2,2]$ which can be represented as $(0,0,2)_{o}$.
Therefore, for the $B_{1u}$-representation, the system shows weak topological crystalline superconductivity protected by $a$-glide symmetry.

Now we discuss the case away from $\mu=0.55$.
Figure~\ref{fig: UCoGe_phase}(c) predicts that topological nodes protected by the glide symmetry emerge with doping holes.
For $-2.0<\mu<-0.8$ or $1.6<\mu<2.0$, incompatibility between the parity of $\theta^a(0)+\theta^a(\pi)$ and the vanishing winding number predicts gapless excitations in the BZ sandwiched by the planes $k_z=0$ and $\pi$.
Direct calculation of the energy spectrum for $\mu=-1.4$ confirms this fact [see Fig.~\ref{fig: Weyl_B3u_B1u}(b)].
As is the case for the $B_{3u}$-representation, classifying paring symmetry provides complimentary understanding which predicts that point nodes emerge along the line $(k_x,k_y)=(0,0)$ (see Appendix~\ref{sec: app_UCoGe_class_pairing_lines}).

%%%
(ii) \textit{$n$-glide symmetry--}
%%%
Table~\ref{table: rep} indicates that the paring potential is even under the $n$-glide transformation.
Thus, the topology protected by $n$-glide symmetry is characterized by the $\mathbb{Z}_2$-indices, $\nu^n_{\pm}(0)$ and $\nu^n_{\pm}(\pi)$.

Figure~\ref{fig: UCoGe_phase}(d) shows that for $\mu=0.55$, the topological indices take $\nu^n_\pm(\pi)=1$ and $\nu^n_\pm(0)=0$, 
meaning that the system is characterized by $(0,1,0)_{e}$.
Therefore, for the $B_{1u}$-representation, the system shows strong topological crystalline superconductivity protected by $n$-glide symmetry.

%%%%%%%%%%%%%%%%%%
\subsubsection{Topology for the $B_{2u}$-representation}
%%%%%%%%%%%%%%%%%%
As seen in Table~\ref{table: rep}, the superconductivity is even for $a$- and $n$-glide transformations. 
Thus, the topology is characterized by the $\mathbb{Z}_2$-indices, $\nu^{a(n)}_{\pm}(0)$ and $\nu^{a(n)}_{\pm}(\pi)$, respectively.
Based on the $K$-theory, it is pointed out that glide-even superconductivity is completely characterized with $\mathbb{Z}_2$-indices~\cite{Yanase_Moebius_UPt3_PRB17}. This result is consistent with the fact that for glide-even superconductivity, the winding number $W_3$ is fixed to zero because glide symmetry plays a similar role to reflection symmetry (see Appendix~\ref{sec: app_UCoGe_R}).

%%%
(i) \textit{$a$-glide symmetry--}
%%%
Figure~\ref{fig: UCoGe_phase}(e) indicates that 
for $\mu=0.55$, the topological indices take $\nu^a_\pm(0)=\nu^a_\pm(\pi)=1$, meaning that the system is characterized by $(0,0,1)_{e}$.
Therefore, for the $B_{2u}$-representation, the system shows weak topological crystalline superconductivity protected by $a$-glide symmetry.

%%%
(ii) \textit{$n$-glide symmetry--}
%%%
Figure~\ref{fig: UCoGe_phase}(f) indicates that for $\mu=0.55$, the topological indices take $\nu^n_\pm(0)=0$, and $\nu^n_\pm(\pi)=1$, meaning that the system is characterized by $(0,1,0)_{e}$. 
Therefore, for the $B_{2u}$-representation, the system shows stong topological crystalline superconductivity protected by $n$-glide symmetry.

%%%%%%%%%%%%%%%%%%%%%%%%
%\input{summary}
%%%%%%%%%%%%%%%%%%%%%%%%
%%%%%%%%%%%%%%%%%%%%%%%%
\section{Summary}
\label{eq: summary}
%%%%%%%%%%%%%%%%%%%%%%%%
%
%In this paper, we have proposed an efficient method to compute the $\mathbb{Z}_4$-indices with glide symmetry for symmetry class AII and DIII. 
In this paper, we have proposed an efficient method to compute the $\mathbb{Z}_4$-indices with glide symmetry which is applicable both for M\"obius topological insulators and M\"obius topological superconductors. 
The former (latter) ones belong to class AII (class DIII), respectively.
The advantage of our method over the existing method based on WCCs is that the former one directly provides the $\mathbb{Z}_4$-indices while the latter one requires careful observations of complicated momentum dependent WCCs. 
This advantage allows us to systematically analyze the topology of systems.

As an application, we have performed systematic computation of the $\mathbb{Z}_4$-index for $\mathrm{CeNiSn}$ which is proposed as a three-dimensional M\"obius Kondo insulator of class AII. The obtained phase diagram basically shows nice agreement with the results obtained by the previous study.
The difference from the previous study is the following. Our efficient method has elucidated that topological index for $-35<\mu_c<-21$ takes the same value as that for $-76<\mu_c<-60$, even though the number of surface Dirac cones is different between these two regions.
This is because the topological state $\theta=1$ is identical to the one with $\theta=-3$ which hosts three Dirac cones at the surface.

Furthermore, we have applied our method to superconductivity of $\mathrm{UCoGe}$ whose topology with glide symmetry had not been fully characterized.
Specifically, our direct computation of $\mathbb{Z}_4$-indices both at the BZ center and at the BZ face has elucidated that $\mathrm{UCoGe}$ shows strong M\"obius superconductivity of the $A_u$- or $B_{3u}$-representation. 
We stress that our numerical method is applicable regardless of the position in the BZ and the additional crystalline symmetry in contrast to the analytic formula proposed in Ref.~\onlinecite{Daido_UCoGe18}.
Thanks to this advantage the $\mathbb{Z}_4$-index for the BZ center is firstly computed without any assumption.
Furthermore, the obtained phase diagrams discover novel gapless excitations in the bulk which are topologically protected by the glide symmetry; the difference of $\mathbb{Z}_4$-indices at the BZ face and the BZ center requires the gapless excitations in the bulk as the difference of the Chern number does for the Weyl superconductivity. 
We have observed this type of gapless excitations with glide symmetry by doping holes into the superconducting phase of the $B_{1u}$- or $B_{3u}$-representation.

%%%%%%%%%%%%%%%%%%%%%%%%
\section{Acknowledgement}
%%%%%%%%%%%%%%%%%%%%%%%%
The authors thank K. Shiozaki and P.-Y. Chang for fruitful discussion.
The authors particularly thank S. Sumita for fruitful comments on the classification of the paring potentials at glide invariant lines.
This work is partly supported by JSPS KAKENHI Grants
No.~JP15K05164, %YY
No.~JP15H05745, %YY
No.~JP15H05855, %NK
No.~JP15H05884, %YY
No.~JP16H00991, %YY
No.~JP16K05501, %NK
No.~JP17J10588, %AD
No.~JP18H01140, %NK
No.~JP18H04225, %YY
No.~JP18H01178, %YY
No.~JP18H05227, %YY
and No.~JP18H05842. %TY
The numerical calculations were performed on the supercomputer at the ISSP in the University of Tokyo and the supercomputer at YITP in Kyoto University.

%%%%%%%%%%%%%%%%%%%%%%%%%%%%%%
%\input{./Master_file_fix.bbl}

%%%%%%%%%%%%%%%%%%%%%%%%
%%%%%%%%%%%%%%%%%%%%%%%%
%%%%%%%%%%%%%%%%%%%%%%%%
\appendix
%%%%%%%%%%%%%%%%%%%%%%%%
%%%%%%%%%%%%%%%%%%%%%%%%
%%%%%%%%%%%%%%%%%%%%%%%%

%%%%%%%%%%%%%%%%%%%%%%%%
%\input{Framework_app.tex}
%%%%%%%%%%%%%%%%%%%%%%%%

%%%%%%%%%%%%%%%%%%
\section{
Integral of the Berry connection for time-reversal symmetric systems
}
\label{sec: AI to A}
%%%%%%%%%%%%%%%%%%
Consider one-dimensional systems with time-reversal symmetry with $\Theta^2=-1$.
In this case, we obtain~\cite{Fu_Z2_PRB06}
%%%%%%%%%%%%%%%%%%
\begin{subequations}
\label{eq: int AI = int A/2}
\begin{eqnarray}
\int^\pi_{-\pi}\!dk_x \mathrm{tr}\mathcal{A}^I(k_x)
&=&
\int^\pi_{-\pi}\!\frac{dk_x}{2} \mathrm{tr}\mathcal{A}(k_x),
\end{eqnarray}
under the gauge choice
\begin{eqnarray}
\label{eq: gauge cont TII}
|u^I_n (-k_x)\rangle    &=&  -\Theta |u^{II}_n (k_x)\rangle, \\
\label{eq: gauge cont TI}
|u^{II}_n (-k_x) \rangle &=&  -\Theta |u^{I}_n (k_x) \rangle,
\end{eqnarray}
\end{subequations}
%%%%%%%%%%%%%%%%%%
for $-\pi \leq -k_x<0$. Superscript $s=I,II$ labels Kramers pairs.
$\mathcal{A}(\bm{k})$ denotes the Berry connection of the occupied bands, $\mathcal{A}(\bm{k})=\mathcal{A}^I(\bm{k})+\mathcal{A}^{II}(\bm{k})$.

In the following we derive the above relation.
With the above gauge choice, we can obtain the following relation:
%%%%%%%%%%%%%%%%%%
\begin{eqnarray}
\int^0_{-\pi}\!dk_x \mathrm{tr}\mathcal{A}^I(k_x) 
&=&\int^{\pi}_0\!dk_x \mathrm{tr}\mathcal{A}^I(-k_x), \nonumber \\
&=&\int^{\pi}_0\!dk_x \mathrm{tr}\mathcal{A}^{II}(k_x).
\end{eqnarray}
%%%%%%%%%%%%%%%%%%
Here we have used the relation: 
%%%%%%%%%%%%%%%%%%
\begin{eqnarray}
\langle u^{I}_n(-k_x)| \partial_{-k_x} | u^{I}_n(-k_x) \rangle 
&=& 
\langle \Theta u^{II}_n (k_x)| (-\partial_{k_x}) | \Theta u^{II}_n (k_x) \rangle 
\nonumber \\
&=& 
\langle \Theta u^{II}_n(k_x)| \Theta (-\partial_{k_y})u^{II}_n(k_x) \rangle
\nonumber \\
&=& 
\langle (-\partial_{k_x})u^{II}_n (k_x) | u^{II}_n(k_x) \rangle
\nonumber \\
&=& 
\langle u^{II}_n(k_x) | \partial_{k_x} u^{II}_n(k_x) \rangle.
\end{eqnarray}
%%%%%%%%%%%%%%%%%%
Therefore, we obtain Eq.~(\ref{eq: int AI = int A/2}).

%%%%%%%%%%%%%%%%%%
\section{
Computing the integral of Berry connection modulo $4\pi$
}
\label{sec: proof mod 4}
%%%%%%%%%%%%%%%%%%
%

Consider a one-dimensional subspace of the two- or three-dimensional BZ $-\pi \leq k < \pi$ where the following two conditions are satisfied: the time-reversal symmetry is closed for this subspace; the Hamiltonian can be block-diagonalized with the glide symmetry.
Let $H_+(k)$ denote the Hamiltonian for plus sector of the glide symmetry whose dimension satisfies $\mathrm{dim}\ H_+ \in 4\mathbb{Z}$.

In this case, we obtain 
%%%%%%%%%%%%%%%%%%
\begin{subequations}
%%%
\begin{eqnarray}
\label{eq: sum A_app}
\sum_{l} A_{+}(k_l) &=& \sum_{l} \tilde{A}_{+}(k_l) +4\pi n,
\end{eqnarray}
%%%
under the following gauge choice: for $-k_l<0$,
%%%
\begin{eqnarray}
\label{eq: gauge_fix_app_latt-mod4_kminus}|
n(-k_l)\rangle &=&- \Theta |n(k_l)\rangle,
\end{eqnarray}
for $k_l=0, \pi$,
\begin{eqnarray}
\label{eq: gauge_fix_app_latt-mod4_TRIM}
|2n+2(k_l)\rangle&=& \Theta |2n+1(k_l)\rangle,
\end{eqnarray}
%%%
\end{subequations}
%%%%%%%%%%%%%%%%%%
where $n$ is an arbitrary integer $n \in \mathbb{Z}$.
$A_{+}(k_l)$ is the lattice version of the Berry connection defined with the occupied states 
$
\psi(\bm{k}):=(|1(\bm{k})\rangle,|2(\bm{k})\rangle,\cdots,|2M(\bm{k})\rangle)
$.
The Berry connection $\tilde{A}_{+}(k_l)$ is computed from $\tilde{\psi}$ which is obtained by applying a unitary matrix to $\psi$ [see Eq.~(\ref{eq: gauge_trans})].
Here, $\sum_l$ takes the summation from $l=0,1\cdots,N-1$. $k_l=-\pi +l(2\pi/N)$. $N$ is an even integer.

Eq.~(\ref{eq: sum A_app}) means that the integration of the Berry connection $A_{+}(k)$ along the one-dimensional line is evaluated modulo 4.
In the following we derive Eq.~(\ref{eq: sum A_app}).

First we define a matrix $M(k_l)$ as 
%%%%%%%%%%%%%%%%%%
\begin{eqnarray}
M(k_l)_{nm} &=& \langle n(k_l)| m(k_{l+1}) \rangle.
\end{eqnarray}
%%%%%%%%%%%%%%%%%%
This matrix satisfies the following relation
%%%%%%%%%%%%%%%%%%
\begin{eqnarray}
\label{eq: appB det M}
\mathrm{det}M(k_l) &=& \mathrm{det}M(k_{N-l-1}),
\end{eqnarray}
for $0\leq l \leq N/2$.

The above relation results in
%%%%%%%%%%%%%%%%%%
\begin{eqnarray}
\label{eq: app_M_half_BZ}
\sum_{l=0,\cdots,N-1}\log \mathrm{det}M(k_l)
&=&
2\sum_{l=N/2,\cdots,N-1}\log \mathrm{det}M(k_l).
\nonumber\\
\end{eqnarray}
%%%%%%%%%%%%%%%%%%

We note that under the gauge transformation $\psi(k_l)_{n}=\tilde{\psi}(k_l)_{m}V_{mn}$, the following relation holds:
%%%%%%%%%%%%%%%%%%
\begin{eqnarray}
\log \mathrm{det}M(k_l)
&&=
 \log \mathrm{det}\tilde{M}(k_l)
-\log \mathrm{det}V(k_l) \nonumber \\
&& \quad \quad 
+\log \mathrm{det}V(k_{l+1})
+2\pi n_0,
\end{eqnarray}
%%%%%%%%%%%%%%%%%%
where the matrix $\tilde{M}(k_l)$ is defined by $\tilde{\psi}$, and $n_0$ takes an integer.

Therefore, by using Eq.~(\ref{eq: app_M_half_BZ}) we obtain
%%%%%%%%%%%%%%%%%%
\begin{eqnarray}
\label{eq: sum M sum M_tilde}
&&
\sum_{l=0,\cdots,N-1}\log \mathrm{det}M(k_l)
\nonumber \\
&&= 
\sum_{l=0,\cdots,N-1}\log \mathrm{det}\tilde{M}(k_l) +4\pi n
\nonumber \\
&&
\quad \quad \quad \quad \quad 
-2[\log \mathrm{det} V(k_{N/2}) -\log \mathrm{det} V(k_{0})],
\end{eqnarray}
%%%%%%%%%%%%%%%%%%
where $n$ is an integer. Because the unitary matrix $V(k_l)$ satisfies
%%%%%%%%%%%%%%%%%%
\begin{eqnarray}
\label{eq: det V=1}
\mathrm{det}V(k_l)&=&1,
\end{eqnarray}
%%%%%%%%%%%%%%%%%%
for $l=0,N/2$, we can see that Eq.~(\ref{eq: sum A_app}) holds.

In the following subsections, we derive Eqs.~(\ref{eq: appB det M})~and~(\ref{eq: det V=1}).

%%%%%%%%%%%%%%%%%%
\subsection{Derivation of Eq.~(\ref{eq: appB det M})}
%%%%%%%%%%%%%%%%%%
For $0<l<N/2$, Eq.~(\ref{eq: appB det M}) can be proven as follows. 
Firstly, we note that the following relation holds because of Eq.~(\ref{eq: gauge_fix_app_latt-mod4_kminus}): for $0<l<N/2$ (i.e., $-\pi<k_l<0$),
%%%%%%%%%%%%%%%%%%
\begin{eqnarray}
\psi(k_l) &=& -U_{\Theta}\psi^*(k_{N-l}),
\end{eqnarray}
%%%%%%%%%%%%%%%%%%
where $U_{\Theta}$ denotes the unitary part of the time-reversal operator satisfying $\Theta^2=-1$, i.e.,  $\Theta:=U_{\Theta}\mathcal{K}$. 
For ordinary time-reversal invariant insulators, $U_{\Theta}$ is given as $U_{\Theta}=is_y\otimes \1$ with the Pauli matrix $s_y$ acting on the spin space and the identity matrix $\1$.
By using the above relation, we have
%%%%%%%%%%%%%%%%%%
\begin{eqnarray}
M(k_l)_{nm}
&=& [\psi^\dagger(k_l)\psi(k_{l+1})]_{nm} \nonumber \\
&=& 
[\psi^T(k_{l+1})\psi^*(k_l)]_{mn}
\nonumber \\
&=& 
[ \{ (-U_{\Theta})\psi^*(k_{N-1-l}) \}^T
\nonumber \\
&&\quad
\{ (-U_{ \Theta }) \psi^*(k_{N-1})\}^*]_{mn}
\nonumber \\
&=&
[\psi^\dagger(k_{N-1-l}) \psi(k_{N-l})]_{mn}
\nonumber \\
&=&
M_{mn}(k_{N-1-l}),
\end{eqnarray}
%%%%%%%%%%%%%%%%%%
which results in Eq.~(\ref{eq: appB det M}) for $0<l<N/2$.

For $l=0,N/2$, Eq.~(\ref{eq: appB det M}) can be proven as follows. Firstly, for $l=0,N/2$, we can find
%%%%%%%%%%%%%%%%%%
\begin{subequations}
\begin{eqnarray}
\label{eq: det M k eq trim help}
[ U_{ \Theta } \psi^*(k_l)]_{j_1m}
&=&
[\psi(k_l)i\tau^y\otimes \1]_{j_1m},
\end{eqnarray}
with $i\tau^y$ transforming the two states as follows:
\begin{eqnarray}
&&
\left(|2n+1\rangle, |2n+2\rangle \right) (i\tau^y) 
\nonumber \\
&&\quad=
\left(
-|2n+2\rangle, |2n+1\rangle
\right),
\end{eqnarray}
\end{subequations}
%%%%%%%%%%%%%%%%%%
with $n=0,1,\cdots, \mathrm{dim}\mathcal{H}_+/4-1$. 
The above relation holds because of the gauge choice~(\ref{eq: gauge_fix_app_latt-mod4_TRIM}).
By making use of Eq.~(\ref{eq: det M k eq trim help}), we can prove Eq.~(\ref{eq: appB det M}) for $l=0$. 
A straightforward calculation yields
%%%%%%%%%%%%%%%%%%
\begin{eqnarray}
M(k_{0})_{n m}&=& 
[\psi^\dagger(k_{0}) 
\psi(k_{1})]_{nm}
\nonumber \\
&=& 
[\psi^T(k_{1})
\psi^*(k_{0})]_{mn}
\nonumber \\
&=& 
[\{ -U_{\Theta}  \psi^*(k_{N-1})\}^T
\nonumber \\
&& \quad \times
\{
 U^\dagger_{ \Theta }\psi(k_{0}) (i\tau^y\otimes\1)
\}]_{mn}
\nonumber \\
&=& 
[
\psi^\dagger(k_{N-1})
\psi(k_{0}) (i\tau^y\otimes\1)
]_{mn}
\nonumber \\
&=& 
[
M(k_{N-1})(i\tau^y\otimes\1)
]_{mn},
\end{eqnarray}
%%%%%%%%%%%%%%%%%%
which results in Eq.~(\ref{eq: appB det M}) for $l=0$.

In a similar way, we obtain
%%%%%%%%%%%%%%%%%%
\begin{eqnarray}
&& M(k_{N/2-1})_{n m} \nonumber \\
&&\quad=
[\psi^\dagger(k_{N/2-1}) 
\psi(k_{N/2})]_{nm}
\nonumber \\
&&\quad=
[
\psi^T(k_{N/2})
\psi^*(k_{N/2-1}) 
]_{mn}
\nonumber \\
&&\quad=
[\{ U_{\Theta} \psi^*(k_{N/2})(-i\tau^y\otimes\1)\}^T
\nonumber \\
&&\quad\quad\quad\quad \times
\{ -U_{\Theta} \psi^*(k_{N/2+1})\}^*]_{mn}
\nonumber \\
&&\quad=
[
(i\tau^y\otimes\1)\psi^\dagger(k_{N/2})\psi(k_{N/2+1})
]_{mn}
\nonumber \\
&&\quad=
[
(i\tau^y\otimes\1)M(k_{N/2})
]_{mn},
\end{eqnarray}
%%%%%%%%%%%%%%%%%%
which results in Eq.~(\ref{eq: appB det M}) for $l=N/2$. We note that $\mathrm{det}(i\tau^y\otimes \1)=1$ holds.
Therefore, we obtain Eq.~(\ref{eq: appB det M}) for $0 \leq  l \leq  N/2$.

%%%%%%%%%%%%%%%%%%
\subsection{Derivation of Eq.~(\ref{eq: det V=1})}
%%%%%%%%%%%%%%%%%%

Let $\{|a_1(k_l)\rangle, |a_2(k_l)\rangle\,\cdots \}$ $(l=0,N/2)$ be a set of eigenstates of the Hamiltonian, spanning $\mathrm{dim}H_+/2$-dimensional Hilbert space.
From this set, $\psi(k_l):=(|1(k_l)\rangle, |2(k_l)\rangle \cdots)$ is obtained as
%%%%%%%%%%%%%%%%%%
\begin{eqnarray}
|2n+1(k_l)\rangle &=& |a_{2n+1}(k_l)\rangle, \nonumber \\
|2n+2(k_l)\rangle &=& \Theta |a_{2n+1}(k_l)\rangle,
\end{eqnarray}
%%%%%%%%%%%%%%%%%%
with $n=0,1,\cdots, \mathrm{dim}\mathcal{H}_+/4-1$.
Here, we consider another set of eigenstates $\{|\tilde{a}_1(k_l)\rangle, |\tilde{a}_2(k_l)\rangle\,\cdots\}$ defined as
%%%%%%%%%%%%%%%%%%
\begin{eqnarray}
(\tilde{a}_1(k_l)\rangle, |\tilde{a}_2(k_l)\rangle\,)
&=&
(|a_1(k_l)\rangle, |a_2(k_l)\rangle\,)U, \nonumber \\
\end{eqnarray}
%%%%%%%%%%%%%%%%%%
with a unitary matrix $U$.
With this new set, one can define $\tilde{\psi}(k_l):=(|\tilde{1}(k_l)\rangle, |\tilde{2}(k_l)\rangle \cdots)$ as
%%%%%%%%%%%%%%%%%%
\begin{eqnarray}
|\widetilde{(2n+1)}(k_l)\rangle &=& |\tilde{a}_{2n+1}(k_l)\rangle, \nonumber \\
|\widetilde{(2n+2)}(k_l)\rangle &=& \Theta |\tilde{a}_{2n+1}(k_l)\rangle.
\end{eqnarray}
%%%%%%%%%%%%%%%%%%
%
These sets of eigenstates of the Hamiltonian, $\psi$ and $\tilde{\psi}$ span the eigenspace of the Hamiltonian and thus, these are related with a unitary matrix,
%%%%%%%%%%%%%%%%%%
\begin{eqnarray}
\tilde{\psi}(k_l)
&=&
\psi(k_l)V(k_l).
\end{eqnarray}
%%%%%%%%%%%%%%%%%%

We here consider Pfaffian of the matrix
%%%%%%%%%%%%%%%%%%
\begin{eqnarray}
w(k_l)
&=&
\psi^\dagger(k_l) (is^y\otimes\1)\psi^*(k_l)
\nonumber \\
&=&i\tau^y\otimes \1,
\end{eqnarray}
%%%%%%%%%%%%%%%%%%
leading to the relation
%%%%%%%%%%%%%%%%%%
\begin{eqnarray}
\mathrm{Pf}w(k_l)
&=&
\mathrm{Pf}\tilde{w}(k_l)=(-1)^{\mathrm{dim}M}.
\end{eqnarray}
%%%%%%%%%%%%%%%%%%
We note the following mathematical formula:
%%%%%%%%%%%%%%%%%%
\begin{eqnarray}
 \mathrm{Pf}(B^TAB)&=&\mathrm{Pf}(A)\mathrm{det}(B),
\end{eqnarray}
%%%%%%%%%%%%%%%%%%
for a skew-symmetric matrix $A$ and an arbitrary matrix $B$.
The above two relations provide Eq.~(\ref{eq: det V=1}), which means that the third and fourth terms of the right hand side of Eq.~(\ref{eq: sum M sum M_tilde}) are zero.

We finish this part with a comment on additional degeneracy at the time-reversal invariant momenta $k_l=0,\pi$. In the presence of additional symmetry, the degeneracy of the energy spectrum can be larger than two. In this case, we need to employ the Gram-Schmidt orthogonalization in order to make $\psi(k_l)$ span the $\mathrm{dim}H_+/4$-dimensional Hilbert space.

%%%%%%%%%%%%%%%%%%%%%%%%%
\section{
Quantization of the index $\theta_3$
}
\label{sec: proof of quantization AII}
%%%%%%%%%%%%%%%%%%%%%%%%%
Here we show that $\theta_3$ defined in Eq.~(\ref{eq: Z4inv_FH_AII}) takes an integer, $\theta_3\in \mathbb{Z}$.

Firstly, we note the following relation holds
%%%%%%%%%%%%%%%%%%
\begin{eqnarray}
\label{eq: app_theta3_logUU_to_logUlogU}
 \log U_1U_2U_3U_4&=& \sum_{i=1,\cdots,4} \log U_i \quad (\mathrm{mod} \ 2\pi),
\end{eqnarray}
%%%%%%%%%%%%%%%%%%
with $U_i \in \mathbb{C}$.
By using this relation, we obtain
%%%%%%%%%%%%%%%%%%
\begin{subequations}
\label{eq: AII_app_stoks}
\begin{eqnarray}
&& \sum_{0\leq k_x \leq \pi,k_z}F_{+zx}(k_x,\pi,k_z) \nonumber \\
&&\quad= -\sum_{k_z} [A_{+z}(\pi,\pi,k_z)-A_{+z}(0,\pi,k_z)] \quad (\mathrm{mod} \ 2\pi), \nonumber \\
&& \\
&& \sum_{0\leq k_x \leq \pi,k_z}F_{+zx}(k_x,0  ,k_z) \nonumber \\
&&\quad= -\sum_{k_z} [A_{+z}(\pi,0  ,k_z)-A_{+z}(0,0  ,k_z)] \quad (\mathrm{mod} \ 2\pi). \nonumber \\
\end{eqnarray}
\end{subequations}
%%%%%%%%%%%%%%%%%%
Thus, Eq.~(\ref{eq: Z4inv_FH_AII}) is rewritten as
%%%%%%%%%%%%%%%%%%
\begin{eqnarray}
\label{eq: Z4inv_FH_AII_app_quantization}
\theta_3&=&\frac{i}{\pi}\sum_{k_z}[A_{+z}(0,\pi,k_z)-A_{+z}(0,0,k_z)] \nonumber \\
&&-\frac{i}{2\pi} \sum_{0\leq k_y \leq \pi,k_z}F_{yz}(0,k_y,k_z) \quad (\mathrm{mod} \ 2).
\end{eqnarray}
%%%%%%%%%%%%%%%%%%
We note that the minus sign in the right hand side of Eqs.~(\ref{eq: AII_app_stoks}a) and (\ref{eq: AII_app_stoks}b) is necessary because we have defined the Berry curvature as Eq.~(\ref{eq: def_latt_A_3D}b).

Applying the time-reversal operator maps states in the Hilbert space labeled with $g_{+}$ to those in the space labeled with $g_{-}$ for $k_x=0$. Namely, for $k_x=0$, a state in the plus sector forms a Kramers pair with the corresponding state in the minus sector.
Keeping this fact in mind, we can see that the right hand side of Eq.~(\ref{eq: Z4inv_FH_AII_app_quantization}) corresponds to the Fu-Kane $\mathbb{Z}_2$-index~\cite{Shiozaki_Moebius_class_PRB16,Fu_Z2_PRB06}.
Therefore, Eq.~(\ref{eq: Z4inv_FH_AII_app_quantization}) takes an integer.

%%%%%%%%%%%%%%%%%%%%%%%%%
\section{
Quantization of the index $\theta$
}
\label{sec: proof of quantization}
%%%%%%%%%%%%%%%%%%%%%%%%%
Here we show that $\theta$ defined in Eq.~(\ref{eq: Z4inv_FH}) takes an integer, $\theta\in \mathbb{Z}$.

By using Eq.~(\ref{eq: app_theta3_logUU_to_logUlogU}), we obtain
%%%%%%%%%%%%%%%%%%
\begin{eqnarray}
\label{eq: proof_of_quantization of Z4DIII}
 \theta &=& \frac{i}{\pi} \sum_{k_y} A_{+y}(0,k_y) \quad (\mathrm{mod}\ 2).
\end{eqnarray}
%%%%%%%%%%%%%%%%%%
%
Here, the summation is taken for $-\pi \leq k_y<\pi$.
We note that the left hand side of the above equation corresponds to the lattice version of $\mathbb{Z}_2$-index for one-dimensional superconductivity of class D~\cite{Alicia_IOP12,Sato_JPSJ17}.
The quantization of this term can be seen as follows. We note that because of particle-hole symmetry, the following relation holds:
%%%%%%%%%%%%%%%%%%
\begin{eqnarray}
\psi^\dagger(-\bm{k}) \psi (-\bm{k}+\Delta \bm{k}_y) &=&\varphi^\dagger(\bm{k}-\Delta \bm{k}_y) \varphi (\bm{k}),
\end{eqnarray}
%%%%%%%%%%%%%%%%%%
where $\varphi (\bm{k}):=C\psi(-\bm{k})$ is the set of eigenvectors for unoccupied states. 
$C$ denotes the particle-hole operator.

Thus, we have 
%%%%%%%%%%%%%%%%%%
\begin{eqnarray}
&&2i \sum_{k_y} A_{+y} (0,k_y)\nonumber \\
&&=-2\mathrm{Im} \sum_{k_x=0,k_y} \log \det[\psi^\dagger(\bm{k})\psi(\bm{k}+\Delta \bm{k}_y)] \nonumber \\
&&=-\mathrm{Im} \sum_{k_x=0,k_y} \left( \log \det[\psi^\dagger(\bm{k})\psi(\bm{k}+\Delta \bm{k}_y)] \right. \nonumber \\
&&\quad\quad\quad\quad\quad\quad \left. + \log \det[\varphi^\dagger(\bm{k}-\Delta \bm{k}_y) \varphi (\bm{k})] \right) \nonumber \\
&&=-\mathrm{Im} \sum_{k_x=0,k_y} \log \det[\Phi^\dagger(\bm{k})\Phi(\bm{k}+\Delta \bm{k}_y)] +2\pi N, \nonumber \\
\end{eqnarray}
%%%%%%%%%%%%%%%%%%
with $N\in \mathbb{Z}$. $\Phi(\bm{k})$ is defined as $\Phi(\bm{k}):=\left( \psi(\bm{k}), \varphi(\bm{k})\right)$ satisfying $\Phi(\bm{k})\Phi^\dagger(\bm{k})=\1$.
The summation in the above equation is taken for the closed path which results in an integer multiple of $2\pi$.

Therefore, we can see quantization of the index $\theta$ to an integer.

%%%%%%%%%%%%%%%%%%
\section{
$\mathbb{Z}_2$-index for glide-even superconductivity
}
\label{sec: app_UCoGe_Z2}
%%%%%%%%%%%%%%%%%%

Consider glide-even superconductivity whose BdG Hamiltonian satisfies Eqs.~(\ref{eq: symm_DIII_Z4}a-\ref{eq: symm_DIII_Z4}c) and the following relations
%%%%%%%%%%%%%%%%%%
\begin{subequations}
\label{eq: symm_DIII_Z2_GE}
\begin{eqnarray}
\Theta G(\bm{k})&=&G(-\bm{k}) \Theta,\\
CG(\bm{k})&=&G(-\bm{k}) C.
\end{eqnarray}
\end{subequations}
%%%%%%%%%%%%%%%%%%
The topological structure of this phase is characterized by the $\mathbb{Z}_2$-index defined as~\cite{Shiozaki_Moebius_class_PRB16}
%%%%%%%%%%%%%%%%%%
\begin{eqnarray}
\nu_{\pm} &=& \frac{i}{\pi} \int^{2\pi}_0\!dk_y  \mathrm{tr} \mathcal{A}^I_{\pm y}(k_x=\pi,k_y) \quad (\mathrm{mod}\ 2),
\end{eqnarray}
%%%%%%%%%%%%%%%%%%
which is identical to the $\mathbb{Z}_2$-index for one-dimensional topological superconductivity of class DIII.
This topological index can be computed in a similar way as the case of $\mathbb{Z}_4$-index.
The lattice version of this index is defined as
%%%%%%%%%%%%%%%%%%
\begin{eqnarray}
\label{eq:Z2_inv_FH_2D}
\nu_{\pm} &=& \frac{i}{2\pi} \sum_{k_y} \mathrm{tr} A_{\pm y}(k_x=\pi,k_y) \quad (\mathrm{mod}\ 2),
\end{eqnarray}
%%%%%%%%%%%%%%%%%%
with the gauge choice, Eq.~(\ref{eq: lattice_gauge_fix}).
By employing the lattice version of the $\mathbb{Z}_2$-index, we have obtained $\nu_{\pm}$ shown in Fig.~\ref{fig: UCoGe_phase}.

When the glide sector preserves the inversion symmetry and the pairing potential is parity-odd, the topological structure is governed by the dispersion relation $\epsilon_\alpha$ of the normal state:
%%%%%%%%%%%%%%%%%%
\begin{eqnarray}
(-1)^{\nu_{\pm}}&=&\prod_\alpha \mathrm{sgn}\epsilon_{2\alpha}(\pi,0)\mathrm{sgn}\epsilon_{2\alpha}(\pi,\pi).
\end{eqnarray}
%%%%%%%%%%%%%%%%%%
Here $(k_x,k_y)=(\pi,0)$ and $(\pi,\pi)$ are time-reversal invariant momenta. At these points, we can observe Kramers degeneracy $\epsilon_{2\alpha}=\epsilon_{2\alpha+1}$. 
The symbol $\prod_{2\alpha}$ denotes taking product for one of each Kramers pair.

%%%%%%%%%%%%%%%%%%%%%%%%
%\input{Appl_CeNiSn_APPENDIX.tex}
%%%%%%%%%%%%%%%%%%%%%%%%
\vspace{1.0cm}
%%%%%%%%%%%%%%%%%%%%%%%%%
\section{
Computation of $\mathbb{Z}_4$-index based on WCCs
}
\label{sec: WCC}
%%%%%%%%%%%%%%%%%%%%%%%%%
In this section, by computing WCCs, we show that the $\mathbb{Z}_4$-index takes $\theta=1$ for the topological Kondo insulator~(\ref{eq: Hami_MoebiusKondo}) at $\mu_c=-30$.
Since the glide symmetry defined in Eq.~(\ref{eq: glide_CeNiSn}) maps $k_z \to -k_z$, the $\mathbb{Z}_4$-index is defined as
%%%%%%%%%%%%%%%%%%
\begin{widetext}
\begin{eqnarray} 
\label{eq: 3D_Z4_WCC} 
\theta_3&=& -\frac{i}{2\pi} \int^{\pi}_0\!dk_z \int^\pi_{-\pi}\!dk_y F_{yz}(0,k_y,k_z) -\frac{i}{\pi} \int^{\pi}_0\!dk_x \int^\pi_{-\pi}\!dk_y F_{+xy} (k_x,k_y,0) +\frac{2i}{\pi} \int^{\pi}_{-\pi}\!dk_y A^{I}_{+y}(\pi,k_y,0) \nonumber \\
        &&   +\frac{i}{\pi} \int^{\pi}_0\!dk_x \int^\pi_{-\pi}\!dk_y F_{+xy} (k_x,k_y,\pi)-\frac{2i}{\pi} \int^{\pi}_{-\pi}\!dk_y A^{I}_{+y}(\pi,k_y,\pi) \quad (\mathrm{mod}\ 4).
\end{eqnarray}
\end{widetext}
%%%%%%%%%%%%%%%%%%
We note that the glide operator does not flip the momentum for $k_z=0,\pi$, and thus the Hamiltonian can be block-diagonalized into the plus and the minus sector of the glide operator. The WCC corresponds to polarization of each orbital and is represented as the integral of the corresponding Berry connection (see below)~\cite{ Soluyanov_WCC_PEB2011_01, Soluyanov_WCC_PEB2011}. Thus, by observing the momentum dependence, we can read off the contribution from integral of the Berry connection and the Berry curvature. The latter can be rewritten with the Berry connection by using the Stokes theorem.

\begin{figure}[!h]
\begin{minipage}{1\hsize}
\begin{center}
\includegraphics[width=\hsize,clip]{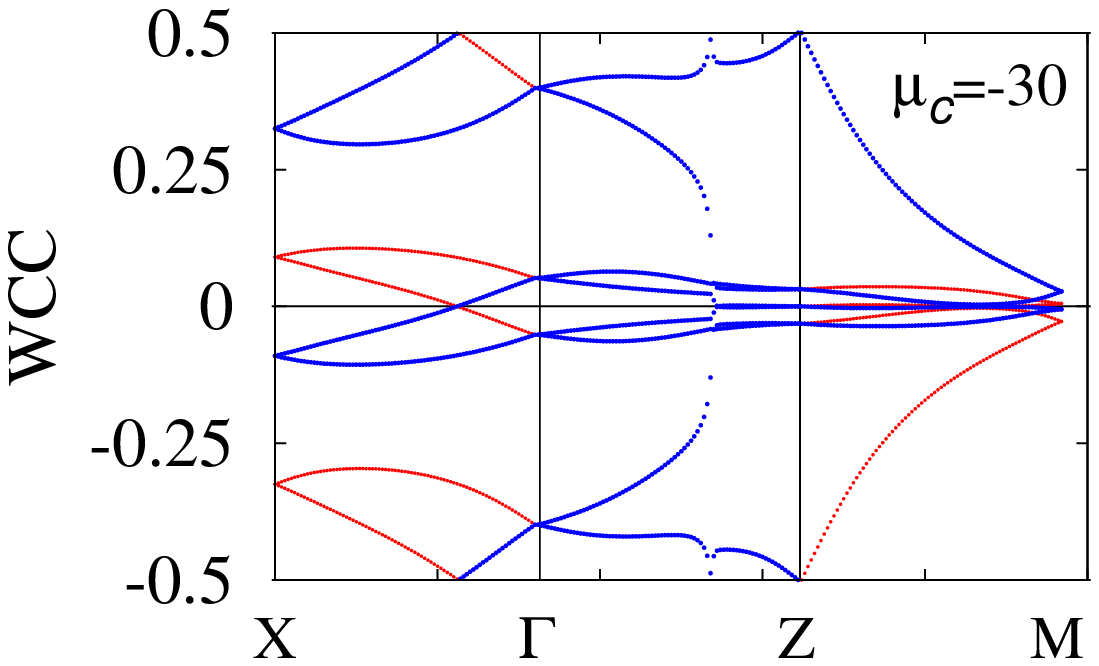}
\end{center}
\end{minipage}
\begin{minipage}{1\hsize}
\begin{center}
\includegraphics[width=\hsize,clip]{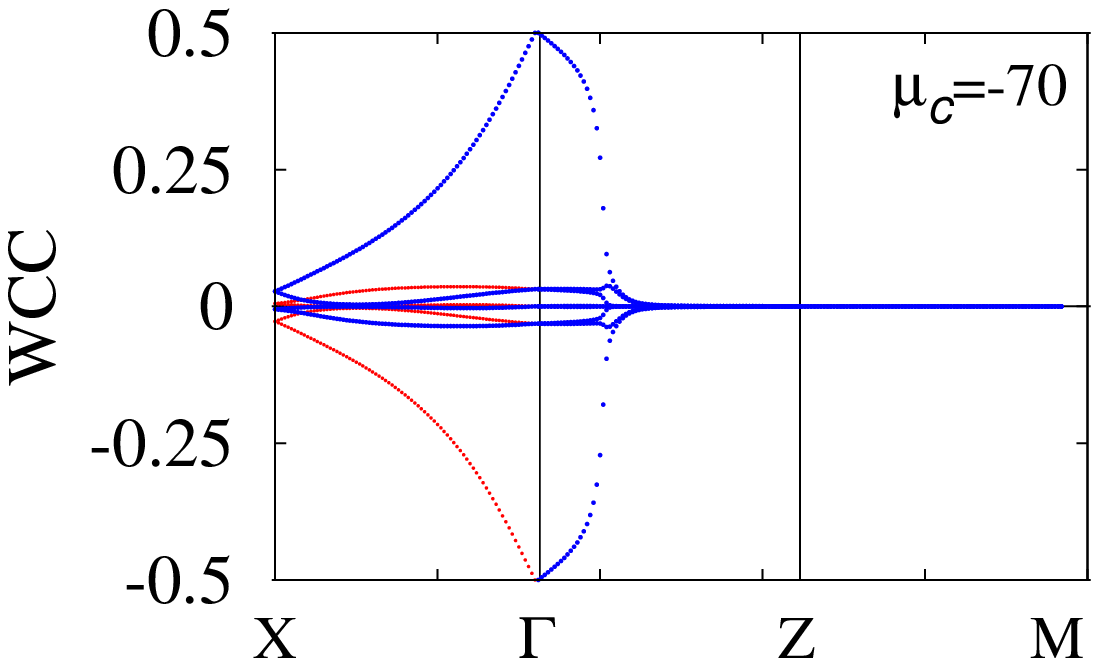}
\end{center}
\end{minipage}
\caption{(Color Online). 
Momentum dependence of the Wannier charge centers (WCCs). Here, X, $\Gamma$, Z, and M points represent $(k_x,k_z)=(\pi,0)$, $(0,0)$, $(0,\pi)$, and $(\pi,\pi)$, respectively.
For $k_z=0,\pi$, the Hamiltonian can be block-diagonalized with the glide symmetry. In these regions, the blue dots represent the data obtained for the plus sector of the glide, while the red dots represent the data obtained for the full Hamiltonian without the block-diagonalization.
These data are obtained for $160\times160\times160$ $k$ mesh. 
For coarser $k$ mesh, the fine structure along the $\Gamma$-Z line cannot be captured.
}
\label{fig: WCCCeNiSn}
\end{figure}

Now, we evaluate the $\mathbb{Z}_4$-index by examining momentum dependence of the WCCs.
In Fig.~\ref{fig: WCCCeNiSn}, the momentum dependence of WCCs is plotted. 
We note that the WCCs in this figure are obtained by integrating the Berry connection along $k_y$-direction~\cite{Soluyanov_WCC_PEB2011}
%%%%%%%%%%%%%%%%%%
\begin{eqnarray}
\bar{y}_n &=& \frac{i}{2\pi} \int^\pi_{-\pi}\!dk_y \langle u_n (\bm{k}) | \partial_{k_y} | u_n (\bm{k}) \rangle.
\end{eqnarray}
%%%%%%%%%%%%%%%%%%
We plot WCCs with blue dots for the plus sector of the Hamiltonian $H_+$ in the region between X and $\Gamma$ and between Z and M. The red dots represent WCCs obtained from the Hamiltonian $H$ including the plus and the minus sector.
By sweeping $k_x$ along $\Gamma \to \mathrm{X}$, we can evaluate the contribution of the second and the third terms of Eq.~(\ref{eq: 3D_Z4_WCC}). Along this path, we can see that one of the WCCs shows a jump from $-0.5$ to $0.5$, resulting in the contribution $-2$ to $\theta_3$.
By sweeping $k_x$ along $\mathrm{Z} \to \mathrm{M}$, we can evaluate the contribution of the fourth and the fifth terms of Eq.~(\ref{eq: 3D_Z4_WCC}). At the $\mathrm{Z}$ point we can see that summing up the WCCs for the plus sector yields $0.5$, resulting in the contribution $-1$ to $\theta_3$.
We can see that the contribution from the other region is zero.
Thus, in total, we obtain $\theta=-3$, which is consistent with $\theta=1$ (mod 4).

%%%%%%%%%%%%%%%%%%%%%%%%%
%%%%%%%%%%%%%%%%%%%%%%%%%

In a similar way, we obtain $\theta=1$ for $\mu_c=-70$. By sweeping $k_x$ along $\Gamma \to \mathrm{X}$, we can evaluate the contribution of the second and the third terms of Eq.~(\ref{eq: 3D_Z4_WCC}). At the $\Gamma$ point, the one of WCCs takes $0.5$, resulting in the contribution of 1. We can see that contribution from the other region is zero.

Therefore, we can conclude that for $\mu_c=-30$ or $-70$, the topological index takes $\theta=1$, which is consistent with the results obtained by the lattice version of the $\mathbb{Z}_4$-index~(\ref{eq: Z4inv_FH_AII}) (see Fig.~\ref{fig: CeNiSn}).

%%%%%%%%%%%%%%%%%%%%%%%%
%\input{Appl_UCoGe_APPENDIX.tex}
%%%%%%%%%%%%%%%%%%%%%%%%

%%%%%%%%%%%%%%%%%%
\section{
Classification of paring potentials at glide invariant lines
}
\label{sec: app_UCoGe_class_pairing_lines}
%%%%%%%%%%%%%%%%%%

In Sec.~\ref{sec: UCoGe_B3u_n-glide}, and \ref{sec: UCoGe_B1u} we have seen the appearance of robust point nodes protected by topological properties. As briefly explained in those sections, one can complementarily understand the emergence of the topological gapless points by analyzing possible paring symmetry of Cooper pairs for a high symmetry line/plane of the BZ. For instance, suppose that the Cooper pairs of the $B_{3u}$-representation are forbidden along a high symmetry line. In this case, the bulk gap, whose paring potential is the $B_{3u}$-representation, should become zero, resulting in robust gapless excitations. 
These gapless modes are nothing but the ones discussed in Sec~\ref{sec: UCoGe_B3u_n-glide}.
%
%
%%%%%%%%%%%%%%%%%%
\begin{table*}[htb]
\begin{center}
\begin{tabular}{cccccccccc} \hline \hline
$(k_x,k_y)$           & $E$ & $2_z$ & $2_y$ & $2_x$ & $I$ & $\sigma^z$ & $\sigma^y$ & $\sigma^x$ & Decomposition \\ \hline 
$(0,0)$               &  $4$ &     $0$ &     $2$ &     $2$ &  $-2$ &          $2$ &          $0$ &          $0$ & $A_g+ A_u+ B_{2u} + B_{3u}$ \\ 
$(\pi,0)$             & $16$ &     $0$ &     $4$ &    $-4$ &  $-4$ &         $-4$ &          $0$ &          $0$ & $A_g+B_{1g}+3B_{2g}+B_{3g}+3A_u + 3B_{1u} +3B_{2u} + B_{3u}$ \\
%                      &    &       &       &       &     &            &            &            & $$\\
$(0,\pi)$             &  $4$ &     $0$ &    $-2$ &     $2$ &  $-2$ &          $2$ &          $4$ &          $0$ & $A_g+B_{1u}+2B_{3u}$\\
$(\pi,\pi)$           &  $4$ &     $4$ &    $-2$ &    $-2$ &  $-2$ &         $-2$ &          $4$ &          $4$ & $A_g+3B_{1u}$\\ \hline\hline 
$(k_y,k_z)$           & $E$ & $2_z$ & $2_y$ & $2_x$ & $I$ & $\sigma^z$ & $\sigma^y$ & $\sigma^x$ & Decomposition \\ \hline 
$(0,0)$               & $4$   &     $2$ &     $2$ &     $0$ & $-2$  &          $0$ &          $0$ &     $2$      &  $A_g+A_u+B_{1u}+B_{2u}$ \\
$(\pi,0)$             & $16$  &     $4$ &    $-4$ &     $0$ & $-4$  &          $0$ &          $0$ &     $-4$     &  $A_g+3B_{1g} + B_{2g} + B_{3g} +3A_{u} +3 B_{1u} + B_{2u} + 3B_{3u}$ \\
$(0,\pi)$             & $16$  &    $-4$ &     $4$ &     $0$ & $-4$  &          $0$ &          $0$ &     $-4$     &  $A_g+B_{1g}+3B_{2g}+B_{3g} +3A_u + B_{1u} + 3B_{2u} + 3B_{3u}$ \\
$(\pi,\pi)$           & $16$  &    $-4$ &    $-4$ &     $0$ & $-4$  &          $0$ &          $0$ &      $4$     &  $A_g+B_{1g}+B_{2g}+3B_{3g} +A_u  +3B_{1u} + 3B_{2u} + 3B_{3u}$ \\ \hline \hline
$(k_z,k_x)$           & $E$ & $2_z$ & $2_y$ & $2_x$ & $I$ & $\sigma^z$ & $\sigma^y$ & $\sigma^x$ & Decomposition \\ \hline 
$(0,0)$               & $4$   &   $2$   &   $0$   &     $2$ & $-2$  &          $0$ &          $2$ &      $0$     & $A_g+A_u+B_{1u}+B_{3u}$ \\
$(\pi,0)$             & $4$   &  $-2$   &   $0$   &     $2$ & $-2$  &          $4$ &          $2$ &      $0$     & $A_g+B_{2u}+2B_{3u}$ \\
$(0,\pi)$             & $4$   &   $2$   &   $0$   &    $-2$ & $-2$  &          $0$ &          $2$ &      $4$     & $A_g+2B_{1u}+B_{2u}$ \\
$(\pi,\pi)$           & $16$  &  $-4$   &   $0$   &    $-4$ & $-4$  &          $0$ &          $4$ &      $0$     & $A_g+B_{1g}+3B_{2g}+B_{3g} +A_u+3B_{1u} + 3B_{2u} + 3 B_{3u} $\\ \hline \hline
\end{tabular}
\caption{
Character table of Cooper pairs $P_{\bm{k}}$ and IR decomposition. 
$E$ denotes the identity transformation. $2_{x(y,z)}$ denotes the two-fold rotation along $x$- ($y$-,$z$-) axis, respectively.
$\sigma^{x}$, $\sigma^{y}$, and $\sigma^{z}$ denote the reflection mapping $(x,y,z)\to (-x,y,z)$, $(x,y,z)\to (x,-y,z)$, and $(x,y,z)\to (x,y,-z)$, respectively. $I$ denotes the inversion.
The reflection and $\pi$-rotation also act on the spin space.
}
\label{table: paring_z-line}
\end{center}
\end{table*}
%%%%%%%%%%%%%%%%%%
%
%

In the following, we see the details by performing classification of possible paring potentials along high symmetry lines of the BZ. The results are summarized in Table~\ref{table: paring_z-line}.

%%%%%%%%%%%%%%%%%%
\subsection{
Classification scheme
}
%%%%%%%%%%%%%%%%%%
The classification of paring potentials along a high symmetry line is carried out in the following steps~\cite{Micklitz_classi_PRL2017,Kobayashi_classi_PRB2018}.
\begin{enumerate}
 \item Obtain the little group $M_{\bm{k}}$ mapping the momentum as $\bm{k}\to g\bm{k}=\bm{k}$ for the high symmetry line.
 \item Calculate the irreducible representation (IR) of the Bloch state $\gamma_{\bm{k}}$ by using Wigner's criterion~\cite{Herring_PR_1937,Wigner_textbook2012group} (see below).
 \item With Mackey-Bradley theorem~\cite{Bradley_textbook2010mathematical} (see below), calculate the character of Cooper pairs $\chi[P_{\bm{k}}]$ for each transformation. Here, $P_{\bm{k}}$ denotes the representation of the Cooper pair composed of Bloch states located at $\bm{k}$ and $-\bm{k}$.
 \item With comparing the set of $\chi[P_{\bm{k}}]$ and the character table of IRs, decompose $P_{\bm{k}}$ into irreducible representations.
\end{enumerate}

Here, we summarize the Wigner's criterion~\cite{Herring_PR_1937,Wigner_textbook2012group}.
Consider a symmetry group defined as $M:=G+a_0G$. Here $G$ is a unitary group. $a_0$ denotes an anti-unitary transformation.
Let $\Delta$ be the IR of $G$. Then, the irreducible representation of $M$ is obtained as follows:
%%%%%%%%%%%%%%%%%%
\begin{subequations}
\begin{eqnarray}
\label{eq: Wigner}
\sum_{a \in M} \chi[\Delta(a^2)]=\sum_{u \in G} \chi[\Delta((a_0u)^2)]=
\left\{
\begin{array}{ll}
  g &\ \ \mathrm{(a)}\\
 -g &\ \  \mathrm{(b)}, \\
  0 &\ \  \mathrm{(c)}
\end{array}
\right.
\nonumber \\
\end{eqnarray}
%%%%%%%%%%%%%%%%%%
where $g$ denotes order of the group $G$. $\chi[\Delta(a)]$ denotes the character (i.e., trace) of the representation.
For the case (a), where $\sum_{a \in M} \chi[\Delta(a^2)]=g$ holds, the representation is written as
%%%%%%%%%%%%%%%%%%
\begin{eqnarray}
\gamma(u)&=&\Delta(u), \\
\gamma(a_0)&=&U.
\end{eqnarray}
%%%%%%%%%%%%%%%%%%
For the case (b) the representation is written as
%%%%%%%%%%%%%%%%%%
\begin{eqnarray}
\gamma(u)&=&
\left(
\begin{array}{cc}
\Delta(u) & 0 \\
0 & \Delta(u)
\end{array}
\right),
\\
\gamma(a_0)&=&
\left(
\begin{array}{cc}
0 & -U \\
U & 0
\end{array}
\right).
\end{eqnarray}
%%%%%%%%%%%%%%%%%%
For the case (c) the representation is written as
%%%%%%%%%%%%%%%%%%
\begin{eqnarray}
\gamma(u)&=&
\left(
\begin{array}{cc}
\Delta(u) & 0 \\
0 & \Delta^*(a_0^{-1}ua_0)
\end{array}
\right),
\\
\gamma(a_0)&=&
\left(
\begin{array}{cc}
0 & \Delta(a^2_0) \\
1 & 0
\end{array}
\right).
\end{eqnarray}
\end{subequations}
%%%%%%%%%%%%%%%%%%
Here, $U$ is a unitary matrix satisfying $\Delta^*(a_0^{-1}ua_0)=U^{-1}\Delta(u)U$.

Mackey-Bradley theorem provides the relation between the character $\chi[P_{\bm{k}} (m)]$ and $\chi[\gamma_{\bm{k}}(m)]$, where $m$ is an element of the little group $M_{\bm{k}}$~\cite{Bradley_textbook2010mathematical}.
Consider the Cooper pair composed of Bloch states located at $\bm{k}$ and $-\bm{k}$.
In this case, the representation $P_{\bm{k}}$ can be written as the anti-symmetrized Kronecker square of induced representation in
%%%%%%%%%%%%%%%%%%
\begin{eqnarray}
\tilde{M}_{\bm{k}}&:=& M_{\bm{k}}+ IM_{\bm{k}}.
\end{eqnarray}
%%%%%%%%%%%%%%%%%%
With Mackey-Bradley theorem, the character of $P_{\bm{k}}$ is obtained as
%%%%%%%%%%%%%%%%%%
\begin{subequations}
\label{eq: M-B theorem}
\begin{eqnarray}
\chi[P_{\bm{k}} (m)]&=&\chi[\gamma_{\bm{k}}(m)]\chi[\gamma_{\bm{k}}(ImI)],\\
\chi[P_{\bm{k}}(Im)]&=&-\chi[\gamma_{\bm{k}}(ImIm)],
\end{eqnarray}
\end{subequations}
%%%%%%%%%%%%%%%%%%
with $m \in M_{\bm{k}}$.

%%%%%%%%%%%%%%%%%%
\subsection{
Classification for the line $(k_y,k_z)=(0,0)$
}
%%%%%%%%%%%%%%%%%%
We apply the above classification scheme for the symmetry group $Pnma$ along a line $(k_y,k_z)=(0,0)$ in the BZ.

Firstly, we obtain the little group $M_{\bm{k}}$. For this line, the little group $M_{\bm{k}}$ is written as
%%%%%%%%%%%%%%%%%%
\begin{subequations}
\begin{eqnarray}
M_{\bm{k}}&=& G_{\bm{k}}+\{ \Theta I | 0 \}G_{\bm{k}}, \\
G_{\bm{k}}&=&
\{E|0\}T
+\{ 2_x | \bm{t}_0 \} T
+\{ \sigma_y | \bm{t}_y \} T \nonumber \\
&&\quad \quad \quad \quad
+\{ \sigma_z | \bm{t}_x+\bm{t}_z \} T. \nonumber\\
\end{eqnarray}
\end{subequations}
%%%%%%%%%%%%%%%%%%
Here we have used the Seitz notation. 
$2_x$ denotes the $\pi$-rotation for the $x$-axis. $\sigma_y(z)$ denotes the reflection for the plane perpendicular to the $y$- ($z$-) axis. 
$E$ is the identity transformation. 
$I$ represents inversion. $\Theta$ denotes the time-reversal transformation for spin-half electrons.
$\bm{t}$'s are vectors of half-translation defined as $\bm{t}_0=(1/2,1/2,1/2)$, $\bm{t}_y=(0,1/2,0)$, and $\bm{t}_z=(0,0,1/2)$.
%%%%%%%%%%%%%%%%%%
\begin{table}[htb]
\begin{center}
\begin{tabular}{ccccccccc} \hline \hline
           & $E$ & $2_z$ & $2_y$ & $2_x$ & $I$ & $\sigma^z$ & $\sigma^y$ & $\sigma^x$  \\ \hline 
$\chi[P_{\bm{k}}]$               &  4 &     0 &     2 &     2 &  -2 &          2 &          0 &          0  \\ \hline \hline 
           & $E$ & $2_z$ & $2_y$ & $2_x$ & $I$ & $\sigma^z$ & $\sigma^y$ & $\sigma^x$  \\ \hline 
$A_g$      &  1  &   1   &  1    &  1    &  1  &     1      &     1      &     1    \\
$B_{1g}$   &  1  &   1   & $-1$  & $-1$  &  1  &     1      &     $-1$   &   $-1$   \\
$B_{2g}$   &  1  & $-1$  & $-1$  &  1    &  1  &    $-1$    &     1      &   $-1$   \\
$B_{3g}$   &  1  & $-1$  &    1  & $-1$  &  1  &    $-1$    &     $-1$   &     1    \\ \hline
$A_u$      &  1  &   1   &  1    &  1    & $-1$&    $-1$    &     $-1$   &   $-1$   \\
$B_{1u}$   &  1  &   1   & $-1$  & $-1$  & $-1$&    $-1$    &       1    &   1      \\
$B_{2u}$   &  1  & $-1$  & $-1$  &  1    & $-1$&       1    &     $-1$   &      1   \\
$B_{3u}$   &  1  & $-1$  &    1  & $-1$  & $-1$&       1    &        1   &   $-1$   \\ \hline\hline
\end{tabular}
\caption{
Character table for $(k_y,k_z)=(0,0)$
(Top panel): character table of Cooper pairs $\chi[P_{\bm{k}}]$.
(Bottom panel):  character table of each IR.
}
\label{table: ky0kz0}
\end{center}
\end{table}
%%%%%%%%%%%%%%%%%%
Secondly, we obtain the IR of the little group $M_{\bm{k}}$.
The IR of $G_{{\bm{k}}}$ is written as
%%%%%%%%%%%%%%%%%%
\begin{subequations}
\begin{eqnarray}
\Delta_{\bm{k}}(E)&=& \rho_0, \\
\Delta_{\bm{k}}(\{ 2_x | \bm{t}_0 \})&=& ie^{ik_x/2}\rho_x, \\
\Delta_{\bm{k}}(\{ \sigma_y | \bm{t}_y \})&=& i\rho_y,\\
\Delta_{\bm{k}}(\{ \sigma_z | \bm{t}_x+\bm{t}_z \})&=& ie^{ik_x/2}\rho_z,
\end{eqnarray}
\end{subequations}
%%%%%%%%%%%%%%%%%%
%
with the Pauli matrices $\rho$'s.
By making use of the Wigner's criterion, we obtain the IR of the little group $M_{\bm{k}}$.
Namely, because 
%%%%%%%%%%%%%%%%%%
\begin{eqnarray}
\sum_{a\in M_{\bm{k}}}\chi[\Delta( (\Theta I u)^2 )]=4,
\end{eqnarray}
%%%%%%%%%%%%%%%%%%
we obtain the representation as 
%%%%%%%%%%%%%%%%%%
\begin{subequations}
\begin{eqnarray}
\gamma_{\bm{k}}(u)&=&\Delta_{\bm{k}}(u), \\
\gamma_{\bm{k}}(\Theta I)&=&\rho_y,
\end{eqnarray}
\end{subequations}
%%%%%%%%%%%%%%%%%%
with $u\in G_{\bm{k}}$.

Thirdly, we calculate the character of $P_{\bm{k}}$.
With Mackey-Bradley theorem [Eq.~(\ref{eq: M-B theorem})], we obtain the character of each symmetry operation summarized as Table~\ref{table: ky0kz0}.

Comparing top and bottom panels of Table~\ref{table: ky0kz0}, we obtain the decomposition: $A_g+A_u+B_{1u}+B_{2u}$. 
Therefore, the paring potential of the $B_{3u}$-representation is forbidden along the line of $(k_y,k_z)=(0,0)$, which results in the gapless nodes along this line in the bulk BZ.

%%%%%%%%%%%%%%%%%%
\section{
Three-dimensional winding number for reflection-even superconductivity
}
\label{sec: app_UCoGe_R}
%%%%%%%%%%%%%%%%%%
%%%%%%%%%%%%%%%%%%
\begin{widetext}
In this section, we show that the three-dimensional winding number is fixed to zero for the reflection-even superconductivity of UCoGe.
In order to see this, let us start with the definition of the winding number.

For class DIII, the following three-dimensional winding number~\cite{Essin_Winding_PRB} characterizes the topology of the systems described by the Hamiltonian $H$:
%%%%%%%%%%%%%%%%%%
\begin{eqnarray}
\label{eq: W_3D_app}
W_{3}&:=&
\frac{\epsilon^{abc}}{3!2(2\pi)^2}
\int_{BZ}\!d^3\bm{k} \mathrm{tr}[\Gamma H^{-1}(\bm{k})\partial_{k_a}H(\bm{k})H^{-1}(\bm{k})\partial_{k_b}H(\bm{k})H^{-1}(\bm{k})\partial_{k_c}H(\bm{k})],
\end{eqnarray}
where $\Gamma$ is defined by product of the time-reversal and the particle-hole operators. $\epsilon^{abc}$ is the anti-symmetric tensor satisfying $\epsilon^{xyz}=1$.
In the case of UCoGe, the Hamiltonian and the matrix $\Gamma$ are written as  $H(\bm{k}):=H_{\mathrm{BdG}}(\bm{k})$ and $\Gamma:=s_y\sigma_0\eta_0\tau_x$. 
[For definition of the time-reversal and the particle-hole operators, see Eqs.~(\ref{eq: UCoGe_TRsymm})~and~(\ref{eq: UCoGe_PHsymm})].

Now, we show that the winding number is fixed to zero for the reflection-even superconductivity of $\mathrm{UCoGe}$. 
First, we note that for reflection-even superconductivity, $U'(R_y,\bm{k})$ is written as $U'(R_y,\bm{k})= U(R_y,\bm{k})\tau_0$ which results in 
%%%%%%%%%%%%%%%%%%
\begin{eqnarray}
[\Gamma, U'(R_y)]  &=& 0.
\end{eqnarray}
%%%%%%%%%%%%%%%%%%
By using this relation, we can see that the three-dimensional winding number $W_{3}$ satisfies
%%%%%%%%%%%%%%%%%%
\begin{eqnarray}
\label{eq: W_3D_app_ref}
W_{3}&=&
\frac{\epsilon^{abc}}{3! 2(2\pi)^2}
\int_{BZ}\!d^3\bm{k} \mathrm{tr}[\Gamma H^{-1}(\bm{k})\partial_{k_a}H(\bm{k})H^{-1}(\bm{k})\partial_{k_b}H(\bm{k})H^{-1}(\bm{k})\partial_{k_c}H(\bm{k})] \nonumber \\
&=&
\frac{\epsilon^{abc}}{3! (2\pi)^2}
\int_{BZ}\!d^3\bm{k} \mathrm{tr}[\Gamma H^{-1}(R_y\bm{k}')\partial_{k_a}H(R_y\bm{k}')H^{-1}(R_y\bm{k}')\partial_{k_b}H(R_y\bm{k}')H^{-1}(R_y\bm{k}')\partial_{k_c}H(R_y\bm{k}')] \nonumber \\
&=&
-
\frac{\epsilon^{abc}}{3! (2\pi)^2}
\int_{BZ}\!d^3\bm{k}' \mathrm{tr}[\Gamma H^{-1}(R_y\bm{k}')\partial_{k'_a}H(R_y\bm{k}')H^{-1}(R_y\bm{k}')\partial_{k'_b}H(R_y\bm{k}')H^{-1}(R_y\bm{k}')\partial_{k'_c}H(R_y\bm{k}')] \nonumber \\
&=&
-
\frac{\epsilon^{abc}}{3! (2\pi)^2}
\int_{BZ}\!d^3\bm{k}' \mathrm{tr}[U'^{-1}(R_y,\bm{k}')\Gamma U'(R_y,\bm{k}') H^{-1}(\bm{k}')\partial_{k'_a}H(\bm{k}')H^{-1}(\bm{k}')\partial_{k'_b}H(\bm{k}')H^{-1}(\bm{k}')\partial_{k'_c}H(\bm{k}')] \nonumber \\
&=&-W_{3},
\end{eqnarray}
%%%%%%%%%%%%%%%%%%
with 
$(\bm{k}:=R_y\bm{k}')$, $\int_{BZ}\!d\bm{k}:=\int^\pi_{-\pi}\!dk_x \int^\pi_{-\pi}\!dk_y \int^\pi_{-\pi}\!dk_z$.
From the third to the fourth line, we have used 
%%%%%%%%%%%%%%%%%%
\begin{subequations}
\begin{eqnarray}
\label{eq: del_k H(Mk)}
&&
\partial_{k'_y}H(R_y\bm{k}')
\nonumber \\
&=&
\partial_{k'_y}U'(\bm{k}')H(\bm{k}')U'^{-1}(\bm{k}')
+U'(\bm{k}')\partial_{k'_y}H(\bm{k}')U'^{-1}(\bm{k}')
+U'(\bm{k}')H(\bm{k}')\partial_{k'_y}U'^{-1}(\bm{k}')
\nonumber \\
&=&
U'(\bm{k}')
\left(
[C_y, H(\bm{k}') ]
+
\partial_{k'_y} H(\bm{k}')
\right)
U'^{-1}(\bm{k}')
,
\end{eqnarray}
%%%%%%%%%%%%%%%%%%
with
%%%%%%%%%%%%%%%%%%
\begin{eqnarray}
U(\bm{k}')&:=&U(R_y,\bm{k}'), \\
C_y &:=& U'^{-1}(\bm{k}')\partial_{k'_y}U'(\bm{k}')=-is_0\sigma_0
\left(
\begin{array}{cc}
0 & 0 \\
0 & 1
\end{array}
\right)_\eta
\tau_0.
\end{eqnarray}
\end{subequations}
%%%%%%%%%%%%%%%%%%
We note that an additional integrand arises from the first term in the last line of Eq.~(\ref{eq: del_k H(Mk)}).
This additional term, however, does not contribute to the winding number 
because the additional term results in boundary terms which are zero due to periodicity of the BZ. This can be seen by the following straightforward calculation.
Up to the prefactor, the additional term is written as
%%%%%%%%%%%%%%%%%%
\begin{eqnarray}
&&
\epsilon^{ybc}\int\!d\bm{k}' 
\mathrm{tr}\left(
\Gamma H^{-1} [C_y,H] H^{-1} \partial_b H H^{-1} \partial_c H
\right)
\nonumber \\
&&
=\epsilon^{ybc}\int\!d\bm{k}' 
\left[
-\mathrm{tr}\left(
\Gamma C_y \partial_b H H^{-1} (\partial_c H) H^{-1} 
\right)
-\mathrm{tr}\left(
\Gamma C_yH^{-1} \partial_b H H^{-1} \partial_c H  
\right)
\right]
\nonumber \\
&&
=\epsilon^{ybc}\int\!d\bm{k}' 
\left[ \mathrm{tr}\left( \Gamma C_y \partial_b H\partial_c H^{-1} \right) +\mathrm{tr}\left( \Gamma C_y \partial_b H^{-1}\partial_c H \right) \right]
\nonumber \\
&&
=\epsilon^{ybc}\int\!d\bm{k}' 
\left[\partial_b \mathrm{tr}\left( \Gamma C_y H\partial_c H^{-1} \right) +\partial_b \mathrm{tr}\left( \Gamma C_y H^{-1}\partial_c H \right) \right]
\nonumber \\
&&=0,
\end{eqnarray}
%%%%%%%%%%%%%%%%%%
with $H:=H(\bm{k}')$. Here we have used the relation $\{\Gamma,H\}=0$.

Therefore, we can see that Eq.~(\ref{eq: W_3D_app_ref}) holds, indicating that $W_3$ is fixed to zero.
In a similar way, we can show that the winding number $W_3$ is fixed to zero for the glide-even superconductivity.

\end{widetext}
%%%%%%%%%%%%%%%%%%


\begin{thebibliography}{60}
\bibitem [{foo()}]{footnote_UCoGe}%
  \BibitemOpen
  \href@noop {} {}\bibinfo {note} {{ 
%
In Ref.~\onlinecite{Daido_UCoGe18}, the authors discussed $\mathbb{Z}_4$-indices at the BZ center based on the assumption that the topology of the $\mathbb{Z}_4$ superconductivity is governed only by the Fermi surface for the normal state.
However, the relation between topology of the Fermi surface for the normal state and the $\mathbb{Z}_4$-index at the BZ center has not been proven
%
  }}\BibitemShut {NoStop}%
\bibitem [{foo()}]{footnote_Z4Weyl}%
  \BibitemOpen
  \href@noop {} {}\bibinfo {note} {{ 
  We note that reflection symmetry also plays an essential role in the emergence of the topological gapless excitations in the bulk
  }}\BibitemShut {NoStop}%
\end{thebibliography}
\end{document}